\documentclass[a4paper,11pt,reqno]{article}

\usepackage{a4wide}
\usepackage{amsmath,amsfonts,amssymb}
\usepackage[english]{babel}
\usepackage{soul}
\usepackage{nicefrac}
\usepackage[mathscr]{euscript}
\usepackage{setspace}
\usepackage[sort&compress,merge,numbers]{natbib}
\usepackage{comment}
\usepackage{enumitem} 
\usepackage{datetime}

\usepackage{mathrsfs}
\usepackage[T1]{fontenc}
\usepackage{mathpazo}
\usepackage{epsfig}

\usepackage[breaklinks=true]{hyperref}

\numberwithin{equation}{section}

\hypersetup{
			colorlinks=true,
			urlcolor=blue,
			citecolor=magenta,
			linkcolor=blue,
     }

\let\oldsqrt\sqrt
\def\sqrt{\mathpalette\DHLhksqrt}

\def\crbig{\\\noalign{\vspace {3mm}}}

\def\DHLhksqrt#1#2{%
\setbox0=\hbox{$#1\oldsqrt{#2\,}$}\dimen0=\ht0
\advance\dimen0-0.2\ht0
\setbox2=\hbox{\vrule height\ht0 depth -\dimen0}%
{\box0\lower0.4pt\box2}}

\def\beq{\begin{equation}}
\def\eeq{\end{equation}}

\def\ov{\overline}

\author{
  \begin{minipage}{.97\linewidth}
    \vspace{1cm}
       \begin{center}
      \begin{small}
        \textbf{Ignatios Antoniadis},$^{1,2}$
      \textbf{Jean-Pierre Derendinger},$^{2,1}$\\
     \textbf{P. Marios Petropoulos}$^3$ and 
      \textbf{Konstantinos Siampos}$^{4,2}$
              \end{small}
    \end{center}
    \vspace{0.5cm}
    \hspace{2.4cm}\begin{minipage}{.7\linewidth}
\begin{center}     {\it \begin{footnotesize}
\hbox{\kern-3.cm\vbox{\vskip0cm
 \begin{itemize}
               \item[$^1$] Laboratoire de Physique Th\'eorique et \\ Hautes Energies - LPTHE\\ 
        Sorbonne Universit\'e, CNRS,\\ 
        4 place Jussieu, 75005 Paris, France
\vskip0.29cm
      \end{itemize}}
\kern-3cm\vbox{
\begin{itemize}
 \item[$^2$] Albert Einstein Center for Fundamental Physics,\\
Institute for Theoretical Physics,\\ 
University of Bern,\\
Sidlerstrasse 5, 3012 Bern, Switzerland
      \end{itemize}\vskip0.05cm
}}
     \end{footnotesize}}
\end{center}
    \end{minipage}
    \vspace{0.5cm}\begin{minipage}{.7\linewidth}
\begin{center}     
{\it \begin{footnotesize}
\hbox{\kern-0.6cm\vbox{\vskip0cm
 \begin{itemize}
               \item[$^3$] Centre de Physique Th\'eorique,\\ 
        Ecole Polytechnique, CNRS UMR 7644,\\ 
        91128 Palaiseau Cedex, France
\vskip0.29cm
      \end{itemize}}
\kern-3cm\vbox{
\begin{itemize}
 \item[$^4$] Theoretical Physics Department,\\
CERN, \\
 1211 Geneva 23, Switzerland
      \end{itemize}\vskip0.05cm
}
}
     \end{footnotesize}}
\end{center}
     \end{minipage}
  \end{minipage}
}

\date{}

\title{\vspace{3.5cm}
\boldmath \begin{Large}
\textbf{All partial breakings in
${\cal N}=2$ supergravity \\ with a single hypermultiplet}
\end{Large} \unboldmath
}

\begin{document}

\begin{titlepage}
\maketitle
\thispagestyle{empty}

\vspace{-14.5cm}
\begin{flushright}CPHT-RR038.062017\\
CERN-TH-2017-227
\end{flushright}
\vspace{12cm}

\begin{center}
\textsc{Abstract}\\
\vspace{1cm}
\begin{minipage}{1.0\linewidth}

We consider partial supersymmetry breaking in ${\cal N}=2$ supergravity coupled to a single vector and 
a single hypermultiplet. This breaking pattern is in principle possible if the quaternion-K\"ahler space of
the hypermultiplet admits (at least) one pair of commuting isometries. For this class of manifolds, 
explicit metrics exist and we analyse a generic electro-magnetic
(dyonic) gauging of the isometries. An example of partial breaking in Minkowski spacetime has been found 
long ago by Ferrara, Girardello and Porrati, 
using the gauging of two translation isometries on $SO(4,1)/SO(4)$. We demonstrate that no other example
of partial breaking of ${\cal N}=2$ supergravity in Minkowski spacetime exists.
We also examine partial-breaking vacua in anti-de Sitter spacetime that are much less constrained and exist 
generically even for electric gaugings.
On $SO(4,1)/SO(4)$, we construct the partially-broken solution and its global limit which is the 
Antoniadis--Partouche--Taylor model.

\end{minipage}
\end{center}


\end{titlepage}

\onehalfspace
\vspace{-1cm}
\begingroup{
\hypersetup{linkcolor=black}
\boldmath
\tableofcontents
\unboldmath
}\endgroup
\noindent\rule{\textwidth}{0.6pt}

\section{Introduction}

Field theories invariant under global or local ${\cal N}=2$ supersymmetry allow very large classes 
of vector, hyper or tensor multiplet
interactions characterized by specific sigma-model geometries. The existence of realizations in which
zero or one supersymmetry remains unbroken at the ground state of the theory is then a relatively vast and
complicated subject which cannot be addressed in theories with more supersymmetries in which the class of allowed matter and gauge couplings is fatally restrictive. 

Consider for instance the simplest global ${\cal N}=2$ Maxwell theory, defined by an arbitrary prepotential
$F(z)$. Since its scalar fields cannot break the $SU(2)_R$ symmetry, a spontaneous breaking to ${\cal N}=1$ is clearly impossible.\footnote{This statement holds with an arbitrary number of Maxwell multiplets.} 
Antoniadis, Partouche and Taylor \cite{APT} (APT) have however invented many years 
ago a realization with partial breaking
in which the $SU(2)_R$ symmetry is violated by electric and magnetic Fayet--Iliopoulos (FI) terms inducing
a nonlinear deformation of the second supersymmetry 
variation of one gaugino, defining it as the (single) goldstino. The ingredients of the model are then 
a {\it non-canonical} holomorphic prepotential and the FI constants.
More recently \cite{ADM}, a similar 
mechanism has been shown to exist for a single hypermultiplet on a specific class of hyper-K\"ahler manifolds 
with a (translational) isometry, using its off-shell single-tensor dual formulation \cite{LR}.

Local ${\cal N}=2$ supersymmetry is more involved in several aspects. Firstly, the supergravity multiplet
includes the graviphoton and electric--magnetic duality in the local super-Maxwell theory is extended and 
powerful \cite{dual}.\footnote{The APT model does not have charged states and is invariant under 
electric--magnetic duality, upon a simultaneous transformation of the FI electric and magnetic constants.}
Partial breaking requires the generation of a massive 
gravitino ${\cal N}=1$ multiplet, with two spin one fields in a $6_\text{B}+6_\text{F}$ (bosonic plus  fermionic degrees of freedom) on-shell content. Fully spontaneous 
partial breaking requires then at least one physical Maxwell multiplet (for the second massive spin one state)
and one hypermultiplet for the $SU(2)_R$ breaking. The minimal case of one hypermultiplet
on the $SO(4,1)/SO(4)$ quaternion-K\"ahler manifold coupled to a single Maxwell multiplet has been studied 
in detail. It was shown that a partial breaking of ${\cal N}=2$ supersymmetry can be realised for a generic prepotential, so that the APT model is obtained in an appropriate rigid globally supersymmetric limit~\cite{FGP2}. A necessary ingredient\,\footnote{To avoid the obstruction described in refs.~\cite{CGP,CGP2}. } is the gauging of ${\cal N}=2$ supergravity along magnetic directions of vector fields, or alternatively a standard electric gauging in a non-prepotential field basis~\cite{FGP}\,\footnote{
{  For earlier work, see \cite{CGP3,CGP4}.}}.

A more general analysis was also performed~\cite{LST1, LST2} in a class of quaternionic manifolds of dimension $4(n+1)$ that are obtained by the so-called C-map from a special K\"ahler manifold of dimension $2n$, 
corresponding to the effective supergravity of the perturbative type II superstring compactified on a Calabi--Yau threefold~\cite{FS}. The special K\"ahler manifold is associated to the scalars of vector multiplets of the mirror theory, while the extra scalar components are the $2n$ Ramond--Ramond fields and the universal hypermultiplet of the string dilaton parametrising for $n=0$ an $SU(2,1)/SU(2)\times U(1)$ space broken to a quaternionic manifold with four isometries upon inclusion of the perturbative (one-loop) corrections~\cite{AMTVoneloop, AADT, ADPS1}. For $n\ne 0$, it was shown that partial breaking can always be realised in either Minkowski or anti-de Sitter (AdS) vacuum by an appropriate choice of the embedding tensor that defines the directions of the gauging~\cite{dWST, embeddingtensor}, which should have again some non-vanishing magnetic component. Finally, for the case of the single universal hypermultiplet $(n=0)$, no Minkowski $\mathcal{N}=1$ vacuum was found.

In this work, we perform a general analysis of the ${\cal N}=2$ partial breaking in supergravity theories containing a single hypermultiplet with two commuting isometries, gauged by the graviphoton and an additional vector multiplet. 
We work in a prepotential frame and use the embedding-tensor formalism \cite{dWST}, for dyonic gaugings of the graviphoton and of the vector multiplet along two commuting isometries of the hypermultiplet manifold. Our goal is to provide a generic treatment for ${\cal N}=1$ Minkowski vacua for arbitrary quaternion-K\"ahler manifolds, special-K\"ahler metrics and dyonic gaugings. In addition, we would like to obtain the APT model \cite{APT} as an off-shell gravity-decoupling limit. A general quaternionic manifold of dimension four with two commuting isometries can be parametrised by the Calderbank--Pedersen (CP) metric~\cite{CP}, where we find a no-go result for ${\cal N}=1$ Minkowski vacua for a general special K\"ahler manifold of the vector multiplet, which seems to be in contradiction with the results obtained for the hyperbolic space $SO(4,1)/SO(4)$. We prove that this contradiction is only apparent because the latter space cannot be written in a CP form, with its torus symmetry identified within the three-dimensional abelian 
sub-algebra of $SO(4,1)$, as a single exception.

The outline of this paper is as follows. In Sec.~\ref{preliminaries}, we present a brief review of the 
matter-coupled ${\cal N}=2$ supergravity. We first present the ungauged case exhibiting the electromagnetic 
duality transformations in the symplectic formalism (\S~\ref{seckin}). In passing, we show that a 
non-prepotential frame can exclusively arise from a magnetic duality transformation of the theory defined by 
the superconformal prepotential $F=-iX^0X^1$. We then summarize the gauging of isometries for the 
hypermultiplet manifold using the embedding-tensor formalism (\S~\ref{secshiftpot}); in particular, we 
exhibit the relation of the scalar potential to the fermion shifts that provide a convenient way to look for partial 
supersymmetry breaking ${\cal N}=1$ vacua. 
In Sec.~\ref{CPmetric}, we make a systematic analysis in the case of one hypermultiplet with two isometries. 
We present the CP metric (\S~\ref{CPmetricformulas}) and compute the fermion shifts upon gauging its 
isometries (\S~\ref{secshift}) proving a no-go theorem for partial breaking in Minkowski space 
(\S~\ref{secpart}). We also show that partial breaking in AdS is generically possible and 
we give an explicit example using a standard electric gauging of two shift isometries in the case of the universal 
dilaton hypermultiplet in type II superstrings compactified on a Calabi--Yau threefold 
(\S~\ref{secpartADS}). We then 
identify an obstruction for bringing the hyperbolic space in CP coordinates that allows partial breaking in 
Minkowski space (\S~\ref{H2CP}). In Sec.~\ref{SO14Partial}, we return to the general analysis 
of the $SO(4,1)/SO(4)$ which is actually the only quaternionic manifold that does not admit a CP metric 
when the two commuting isometries are shifts in the Poincar\'e coordinates (\S~\ref{secSO4}). We
 construct explicitly the partial breaking Minkowski vacuum and study its off-shell gravity-decoupling limit 
(\S~\ref{secSO4Mink}), as well as non-supersymmetric 
Minkowski vacua (\S~\ref{secSO4N=0}). Section~\ref{Sec5} contains some concluding remarks. 
Finally, we include four appendices. Appendix~\ref{appendixquaternionic} contains useful formulae for the
 gauging of quaternionic manifolds with isometries, 
App.~\ref{appendixhyperbolic} elaborates the hyperbolic space in CP coordinates, App.~\ref{secWard} discusses coordinate 
transformations used to derive the CP metric
and App.~\ref{appproof} proves a result on ${\cal N} = 0$ vacua of the $SO(4, 1)/SO(4)$ model stated in 
\S~\ref{secSO4N=0}.

\boldmath
\section{Matter-coupled ${\cal N}=2$ supergravities}\label{preliminaries}
\unboldmath

\subsection{The kinetic terms}\label{seckin}

The ${\cal N}=2$ target space ${\cal M}$ describing the scalar-field kinetic 
terms of a single hypermultiplet and $n_{\text{V}}$ vector multiplets is factorized,
\begin{equation}
{\cal M}={\cal M}_{\text{H}}\times{\cal M}_{\text{V}}\,,
\end{equation}
and both metrics only depend of the scalar fields of their respective multiplets, a property which by supersymmetry extends to all kinetic terms. 
The hypermultiplet scalar dynamics is encoded in the four-dimensional quaternion-K\"ahler metric ${\cal M}_{\text{H}}$ 
with coordinates $q^u=(q^1,q^2,q^3,q^4)$\,\footnote{We use hypermultiplet scalars and metric with dimension
mass$^0$.}
\begin{equation}
\label{kinetichyper}
{\cal L}_{\text{hyper}}= -\frac{e}{2\kappa^2}\, g^{\mu\nu} \, h_{uv}\, \partial_\mu q^u\partial_\nu q^v .
\end{equation}
A generic quaternion-K\"ahler manifold for $n_{\text{H}}$ hypermultiplets is Einstein with holonomy 
$Sp(2n_{\text{H}})\times SU(2)$,\footnote{Or $G\times SU(2)$, $G\subset Sp(2n_{\text{H}})$.} and dimension $4n_{\text{H}}$. For $n_{\text{H}}=1$, since $Sp(2)\times SU(2) \sim SO(4)$,
a particular characterization is needed: the quaternion-K\"ahler metric is Einstein with an
(anti-) selfdual Weyl curvature tensor. As we will see in the next sections, four-dimensional metrics with 
these properties have been studied 
quite extensively when they admit one or several continuous isometries, which is the case of interest here. 

The ${\cal N}=2$ Maxwell sector is conveniently constructed in the superconformal formulation: 
it is then defined in terms of a prepotential $F(X^I)$ of
$n_{\text{V}}+1$ complex scalar fields, with Weyl weight one. The index $I=0,\ldots, n_{\text{V}}$ includes a compensating 
multiplet. Its component fields include the propagating graviphoton, while its two gauginos and 
complex scalar are  used to gauge-fix superconformal symmetries and solve field equations of auxiliary 
fields in the Weyl multiplet.\footnote{These appear linearly in the lagrangian and then impose constraints.} 
It is a common but unnecessary choice to set $I=0$ as the compensator direction.  Superconformal invariance requires that $F(X^I)$ has Weyl weight two:
\begin{equation}
F\left(X^I\right) = \left(X^0\right)^2 \, F\left(\frac{X^I}{X^0}\right) = \left(X^0\right)^2 \, F(1, z^a) = -i\, \left(X^0\right)^2 \, f(z^a), \qquad a=1,\ldots,n_{\text{V}},
\end{equation}
and $f(z^a)$ is an arbitrary function of the zero-weight scalar fields $z^a=X^a/X^0$ 
in the $n_{\text{V}}$ physical Maxwell multiplets: the Poincar\'e theory is formulated in terms of the scalars
$z^a$.

There is however a subtlety: electric--magnetic duality acts in the Maxwell sector as 
$Sp(2(n_{\text{V}}+1),\mathbb{R})$ linear transformations of the vector of sections 
\begin{equation}
V = 
\begin{pmatrix} 
X^I \\ F_I 
\end{pmatrix}.
\end{equation}
Choosing the section vector $V$ (as a function of a given set of scalar fields) defines a {\it symplectic frame}: 
electric--magnetic duality would imply that the (ungauged, abelian) theory can be equivalently formulated in each
symplectic frame obtained by the action of $Sp(2(n_{\text{V}}+1),\mathbb{R})$ on $V$.\footnote{The symplectic 
orbit of $V$.}
In a {\it prepotential symplectic frame}, there exists $F(X)$ such that the sections are
\begin{equation}
\label{Fframe}
X^I = (X^0, X^0 z^a), \quad
F_I = {\partial\over\partial X^I} F(X), \quad F_0 = -iX^0 [ 2f(z) - z^a f_a ], \quad
F_a = -i X^0 f_a,
\end{equation}
where $f_a= {\partial\over\partial z^a}f(z)$. In a prepotential frame, the symplectic-invariant product
$i( X^I \ov F_I - F_I\ov X^I)$ reads
\begin{equation}
- X^0\ov X^0 [ 2(f+\ov f) - (z^a-\ov z^a )(f_a-\ov f_a)] = - X^0\ov X^0 \, {\cal Y}
\end{equation}
and ${\cal Y}$ will appear in the K\"ahler potential of the Poincar\'e fields $z^a$.
Note that there is an ambiguity: this quantity vanishes if 
\begin{equation}
\widehat F(X^I) = \alpha_{IJ} X^I X^J, \qquad\qquad \widehat f(z^a) = i[\alpha_{00}
+ 2\alpha_{0a} z^a + \alpha_{ab} z^az^b]
\end{equation}
with real coefficients $\alpha_{IJ}$ and two prepotentials differing by $\widehat F$ describe the same theory.

One may wonder if all frames in the symplectic orbit of a 
prepotential frame admit a prepotential, or if there exists orbits which relate prepotential and 
{\it non-prepotential} frames. The question of the existence of a prepotential frame has been discussed 
in general in Ref.~\cite{CRTVP}\footnote{Summarized in \cite{FVP}, \S~21.2.2, page 474.
See also ref.~\cite{Ceresole:1995jg}.}, but the 
simple case $n_{\text{V}}=1$, which is of interest here is very simple to solve explicitly. 

Consider a symplectic transformation relating sections $V$ and $\widetilde V$, assuming that $V$ defines
a prepotential frame with prepotential $F(X^0, X^1)$ and Poincar\'e scalar $z=X^0/X^1$. Assuming that
we identify the compensators in both frames, $X^0=\widetilde X^0$, $Sp(2,2,\mathbb{R})$
duality reduces to $Sl(2,\mathbb{R})$ transformations
\begin{equation}
\widetilde X^1 = m_1 X^1 + m_2F_1, \qquad \widetilde F_0=F_0, \qquad
\widetilde F_1 = m_3 X^1 + m_4F_1, \qquad (m_1m_4-m_2m_3=1),
\end{equation}
which are electric--magnetic if $m_2\ne0$ or $m_3\ne0$.
We wish to find a Poincar\' e scalar $\widetilde z = \widetilde X^1/\widetilde X^0$ and a prepotential 
$\widetilde F(\widetilde X^I) = -i(\widetilde X^0)^2 g(\widetilde z)$, which identify sections $\widetilde V$ as a prepotential frame:
\begin{equation}
\left\{\begin{array}{l}
\widetilde X_1 = \widetilde X^0 \widetilde z = X^0(m_1 z - i m_2 f_z) ,
\crbig
\widetilde F_1 = -i\widetilde X^0 g_{\widetilde z} = X^0(m_3 z - i m_4 f_z) ,
\crbig
\widetilde F_0 = -i\widetilde X^0 [ 2g(\widetilde z) - \widetilde z g_{\widetilde z} ]
= -iX^0 [ 2f(z) - z f_z ],
\end{array}\right.
\quad\Longrightarrow\quad
\left\{\begin{array}{l}
\widetilde z = m_1 z -im_2f_z,
\crbig
z = m_4\widetilde z + i m_2 g_{\widetilde z},
\crbig
2g(\widetilde z) - \widetilde z g_{\widetilde z} = 2f(z) - z f_z.
\end{array}\right.
\end{equation}
The three equations relating $f$ and $z$ with $g$ and $\widetilde z$ are generated by the Legendre transformation 
\begin{equation}
m_2 g(\widetilde z) - {i\over2}m_4\widetilde z^2 = -i z\widetilde z + m_2f(z) + {i\over2}m_1z^2,
\end{equation}
which exchanges $z$ and $\widetilde z$.
Clearly, the terms induced by $m_1$ or $m_4$ are irrelevant: they modify $f(z)$ or $g(\widetilde z)$
by quadratic terms with imaginary coefficient which do not contribute to the theory. The only relevant 
case is then $m_2= -m_3^{-1}$. The Legendre transformation implies 
\begin{equation}
m_2^2 \, g_{\widetilde z\widetilde z} \, f_{zz} = 1,
\end{equation}
and it is singular only if $f(z)$ (or $g(\widetilde z))$ is linear.  
Hence, the symplectic frame with sections $\widetilde V$ is a prepotential frame with Poincar\'e field
$\widetilde z$ and prepotential $\widetilde F = -i(\widetilde X^0)^2 g \left(\frac{\widetilde X^1}{\widetilde X^0}\right)$ with a single 
exception,
\begin{equation}
\label{Flinear}
F(X^I) = - i \alpha X^0X^1, \qquad\qquad f(z)=\alpha z \qquad\qquad \makebox{($\alpha$ real)},
\end{equation}
for which (with $m_1=m_4=0$)
\begin{equation}
\label{exc}
\widetilde X^0 = X^0, \qquad
\widetilde X^1 = -i\alpha m_2 X^0, \qquad
\widetilde F_0 = -i\alpha X^1 = -i\alpha X^0 z, \qquad
\widetilde F_1 = -m_2^{-1}X^1,
\end{equation}
and 
\begin{equation}
\label{exc2}
i( X^I \ov F_I - F_I\ov X^I) = -\alpha X_0\ov X_0 (z+\ov z)
\end{equation}
leading to K\"ahler potential ${\cal K}=- \ln(z+\ov z)$.
This simple discussion agrees with the general argument given in Refs.~\cite{CRTVP} and 
\cite{FVP}.\footnote{But disagrees with statements in Ref.~\cite{FGP2} for instance. }
The conclusion is that in the $n_{\text{V}}=1$ case, all symplectic orbits connect exclusively prepotential
frames, with the single exception of the orbit of $F(X)=-iX^0X^1$ which includes non-prepotential frames. 

The first example of partial ${\cal N}=2$ breaking in supergravity \cite{FGP} was found using
precisely the sections (\ref{exc}). An electric gauging of two translation isometries of the 
hypermultiplet manifold $SO(4,1)/SO(4) \sim Sp(2,2) / SU(2)\times SU(2)$ in this non-prepotential frame
leads to a two-coupling theory with zero potential and ${\cal N}=0$ for generic values of the couplings,
${\cal N}=1$ when a linear relation is verified by the couplings, and ${\cal N}=2$ for zero couplings.

Since the prepotential (\ref{Flinear}) is in the symplectic orbit of the non-prepotential frame (\ref{exc})
and since all other orbits include prepotential frames only, we are always allowed to work in a prepotential
frame with sections (\ref{Fframe}) and to gauge isometries in this frame: since gauging fixes the electric--magnetic duality symmetry, the theory will then depend on the prepotential, the gauge couplings and the choice of hypermultiplet manifold.

The kinetic terms of the helicity $0, \pm\frac{1}{2}$ fields\footnote{Propagating or auxiliary.} 
in Poincar\' e Maxwell multiplets have a K\"ahler metric with K\"ahler potential
\begin{equation}
\label{Kis}
{\cal K} = - \ln \left[ 2(f+\ov f) - (z^a-\ov z^a )(f_a-\ov f_a)\right].
\end{equation}
For instance, for scalar fields (in a prepotential frame), the superconformal lagrangian includes
\begin{equation}
e^{-1}\,{\cal L}_{\text{kin.}}  = - g^{\mu\nu} N_{IJ} \left(D_\mu X^I\right) \left(D_\nu\ov X^J\right),  
\end{equation}
with
\begin{equation}
N_{IJ} = -iF_{IJ}+i \ov F_{IJ} = {\partial^2N \over\partial X^I \partial \ov X^J}, \qquad\qquad
N = -i X^I(F_{IJ}-\ov F_{IJ})\ov X^J = i \left(X^I \ov F_I - \ov X^I F_I\right),
\end{equation}
and with a  covariant derivative $D_\mu X^I = (\partial_\mu -iA_\mu) X^I$ involving the gauge field of the 
superconformal $U(1)_{R}$  symmetry. Eliminating this auxiliary vector field delivers\,\footnote{Omitting 
fermions.}
\begin{equation}
e^{-1}\,{\cal L}_{\text{kin.}}  =\displaystyle - N \left[{1\over4}\,(\partial_\mu \ln N)(\partial^\mu \ln N)
+ \, {\partial^2\ln N \over \partial X^I\partial \ov X^J}\left(\partial_\mu X^I\right)\left(\partial^\mu\ov X^J\right) \right],
\end{equation}
using the homogeneity of the prepotential. The Poincar\'e theory can then be obtained in field coordinates 
$X^I = X^0(1,z^a)$, $X^0= \kappa^{-1} y(z,\ov z)$ and sections $V =  y U = y(Z^I(z), F_I(z))$ 
once the dilatation and $U(1)_{R}$ gauge-fixing conditions 
\begin{equation}
N= -\kappa^{-2} \quad\longrightarrow\quad (y\ov y)^{-1} =  2(f+\ov f) - (z^a-\ov z^a)(f_a-\ov f_a)
= {\cal Y},
\qquad\quad y=\ov y
\end{equation}
have been applied. In terms of $z^a$ then, 
\begin{equation}
e^{-1}\, {\cal L}_{\text{kin.}}  =
- {1\over\kappa^2} \, g_{a\ov b}\, (\partial_\mu z^a) (\partial^\mu\ov z^b) , \qquad
g_{a\ov b} = {\partial^2{\cal K} \over \partial z^a \partial\ov z^b}, \qquad 
{\cal K}=-\ln{\cal Y}, \qquad
y(z,\ov z) = \text{e}^{{\cal K}/2},
\end{equation}
which leads to expression (\ref{Kis}).
The same metric appears in the kinetic terms of the Poincar\' e gauginos $\lambda^{ia}$.
However the kinetic terms of gauge fields include further contributions due to the graviphoton:
\begin{equation}
e^{-1}\,{\cal L}_{\text{gauge}}  =\frac{\text{1}}{4}\,\text{Im}\,{\cal N}_{IJ\,}F_{\mu\nu}^I\,F^{\mu\nu J}-
\frac{e}{8}\,\text{Re}\,{\cal N}_{IJ}\,\varepsilon_{\mu\nu\rho\sigma}\,F^{\mu\nu I}\,F^{\rho\sigma J}\,,\end{equation}
with $n_{\text{V}}+1$--dimensional metric
\begin{equation}
{\cal N}_{IJ}=\ov F_{IJ}+i\,\frac{N_{IK}X^KN_{JL}X^L}{N_{MN}X^M X^N}\,, \qquad\qquad
I,J=0,\ldots,n_{\text{V}}.
\end{equation} 
Notice that ${\rm Im}{\cal N}_{IJ}$ is negative on physical fields. 

In the following, we will explicitly consider the case $n_{\text{V}}=1$ only.

\subsection{Fermion shifts, scalar potential, supersymmetry breaking}\label{secshiftpot}

In ${\cal N}=2$ supergravity, the scalar potential appears when isometries of the theory are gauged. 
With the graviphoton and the gauge field of a vector multiplet ($n_{\text{V}}=1)$, we can gauge two commuting 
isometries, as required if partial supersymmetry breaking is envisaged \cite{LST1}. This of course implies that two commuting
isometries should exist and this defines a class of scalar manifolds for a single hypermultiplet for which explicit 
metrics are available. The problem of partial breaking can then be analytically studied in general.

The scalar potential in supergravity theories has a particular structure. The supersymmetry variation 
of all fermions $\psi_I^A$ is of the form
\begin{equation}
\delta\, \psi_I^A \sim {\cal M}_{Ij}^A \, \epsilon^j + \cdots,
\end{equation}
where the {\it fermion shift} ${\cal M}_{Ij}^A $ is a function of scalar fields ($A$ runs over all supermultiplets, 
$I$ over all fermions in multiplet $A$, $j$ over all supersymmetries; 
in $\mathcal{N}=2$ theories fermions are always in $SU(2)$ doublets and $I=i$). 
If the supermultiplet admits an off-shell realization, as the ${\cal N}=2$ Maxwell or single-tensor multiplets, 
the fermion shifts are in general auxiliary scalar fields. For instance, in a Maxwell ${\cal N}=2$ multiplet
with gauginos $\lambda^i$,
\begin{equation}
\delta\, \lambda^i \sim Y^{ij} \, \epsilon_j + \cdots, \qquad\qquad Y^{ij}=Y^{ji},
\end{equation}
and $Y^{ij}$ is the $SU(2)$ triplet of real (electric) auxiliary fields.
For the gravitinos,
\begin{equation}
\delta\,\psi_\mu^i \sim {1\over2}\,\kappa^2\, S^{ij}\,\gamma_\mu\epsilon_j + \cdots.
\end{equation}
The scalar potential is then symbolically \cite{FM}
\begin{equation}
\mathscr{V}= e \sum\, \text{coeff.} \times \text{fermion shifts}^\dagger\,\times\text{metric}\times\, \text{fermion shifts},
\end{equation}
where the sum is over all fermions and the coefficients are negative for gravitinos and positive for spin-$\frac{1}{2}$ fields
and depend on the normalization chosen for the fermion fields.
Hence, fermion shifts define the ground state of the theory and a nonzero value of a spin-$\frac{1}{2}$ fermion shift
at the ground state indicates the presence of a goldstino, or several goldstinos, and then indicates spontaneous supersymmetry 
breaking. Analyzing the structure of the fermion shifts is fundamental when studying the breaking phases of a supersymmetric theory.

In order to obtain the fermion shifts, we need to specify the gauging applied in the theory. 
The gauge generators (associated with gauge field $I$) and the gauge variations can be defined by 
electric--magnetic symplectic vectors $\Theta_I{}^a\xi_a$: the embedding tensor $\Theta_I{}^a$ specifies 
a linear combination of the (commuting) isometries $\xi_a=\xi^u_a\,\partial_u$ of the 
quaternion-K\"ahler metric $h_{uv}$; it defines the coupling constants of the gauged theory. The index $I$
defines $\Theta_I{}^a$ as a fixed symplectic vector associated with each isometry, but we will rather use 
\begin{equation}
\Theta_I{}^a = \Omega_{IJ} g^{Ja},
\end{equation}
and the coupling constants are the numbers $g^{Ia}$.

Consistency of the gauging is guaranteed by the locality constraint on the embedding tensor \cite{dWST}
\begin{equation}
\label{embeddingconsistency}
\Theta_I{}^a \Omega^{IJ}\Theta_J{}^b=0\,, \qquad\qquad
\Omega = \left(\begin{array}{cc} 0 & \mathbb{I} \\ -\mathbb{I} & 0 \end{array}\right).
\end{equation}

The hypermultiplet scalar fields are coordinates $q^u$ on a (four-dimensional) quaternion-K\"ahler space with
metric $h_{uv}$. For each isometry vector $\delta_a q^u = \xi^u_a$, one can derive an $SU(2)$ triplet of 
prepotentials (or moment maps) solving the differential equation
\begin{equation}
\label{Killing.prepotential}
P^x_a=-\frac{1}{2\kappa^2}\left(J^x\right)^u{}_v\nabla_u\xi_a^v\,, \qquad\qquad x=1,2,3
\end{equation}
in terms of the triplet of complex structures $J^x$.\,\footnote{Our $SU(2)$ conventions are as in \cite{FVP} -- see also App.~\ref{appendixquaternionic}.}
As usual, to describe the hypermultiplet fermions (hyperinos), we need a vielbein $f^{iA}{}_u$, which for 
${\cal N}=2$ is defined by
\begin{equation}
\label{metric.symplectic}
f^{iA}{}_u\Omega_{AB}f^{jB}{}_v=
\frac{i}{2}(J_x)_{uv} (i\varepsilon\,\sigma^x)^{ij}+\frac12\,h_{uv}\, \varepsilon^{ij}
\quad\Longrightarrow\quad
h_{uv}=f^{iA}{}_u\,\varepsilon_{ij}\Omega_{AB}f^{jB}{}_v
\end{equation}
($i$ and $A$ are respectively $SU(2)$ and $Sp(2n_{\text{H}})=Sp(2)$ doublet indices and hyperinos carry index $A$).
Then, for given quaternion-K\"ahler metric $h_{uv}$, complex structures $J^x$, isometries $\xi^u_a$ and 
prepotentials
\begin{equation}
P^{ij}_a = P^x_a ( i\varepsilon\,\sigma^x)^{ij} = P^{ji}_a, 
\end{equation}
we obtain the following expressions for the fermion shifts:\,\footnote{These shifts hold for fermions with
dimension mass$^{\frac{1}{2}}$. Similarly, the scalars and the metrics $g_{\alpha\ov\beta}$ and 
$h_{uv}$ are dimensionless.
We use Weyl spinors, $\psi_\mu^i$, $\lambda^\alpha_i$, $\zeta^A$, $\epsilon^i$ are left-handed, 
$\psi_{\mu i}$, $\lambda^{\alpha i}$, $\zeta_A$, $\epsilon_i$ are right-handed. The
$SU(2)$ indices are moved with
$\lambda^i=\varepsilon^{ij}\lambda_j$ and $\lambda_i=\lambda^j\varepsilon_{ji}$. } 
\begin{equation}
\label{fermion}
\begin{array}{lll}
\makebox{Gravitinos:}\qquad\quad& \displaystyle 
S^{ij}= {1\over\kappa}\text{e}^{{\cal K}/2} P^{ij}_a U^I \, \Theta_I{}^a   = S^{ji}\,,
& \displaystyle \delta\, \psi_\mu^i = {1\over2}\kappa^2S^{ij}\gamma_\mu\epsilon_j + \cdots, 
\crbig
\makebox{Gauginos:}\qquad\quad&\displaystyle
W_\alpha{}^{ij}= - {1\over\kappa}\,\text{e}^{{\cal K}/2} P_a^{ij} \nabla_\alpha U^I \, 
\Theta_I{}^a 
= W^{ji}_\alpha\,, \qquad
& \delta\, \lambda^\alpha_i  = \kappa^2\,g^{\alpha\ov\beta} \ov W_{\ov\beta ij}\epsilon^j + \cdots,
\crbig
\makebox{Hyperinos:}\qquad\quad&\displaystyle
N^i{}_A= {i\over\kappa}\, \text{e}^{{\cal K}/2} f^{iB}{}_u U^I \, \Theta_I{}^a\xi^u_a \,
\Omega_{BA}, 
& \delta\, \zeta^A = \ov N_i{}^A \epsilon^i + \cdots,
\end{array}
\end{equation}
and of their conjugates ($P_{aij}= P^{ij}_a{}^*$):
\begin{equation}
\begin{array}{ll}
\makebox{Gravitinos:}\qquad\qquad&\displaystyle
\ov S_{ij}=  {1\over\kappa}\text{e}^{{\cal K}/2}P_{aij} \ov U^I\,\Theta_I{}^a \,,\crbig
\makebox{Gauginos:}\qquad&\displaystyle
\ov W_{\ov\alpha ij}= - {1\over\kappa}\, \text{e}^{{\cal K}/2} P_{aij}\ov\nabla_\alpha\ov U^I\,
\Theta_I{}^a\,,
\crbig
\makebox{Hyperinos:}\qquad&\displaystyle
\ov N_i{}^A=- {i\over\kappa}\,\text{e}^{{\cal K}/2}f^{jA}{}_u \ov U^I\, \varepsilon_{ij} 
\Theta_I{}^a\xi^u_a \,.
\end{array}
\end{equation}
The embedding tensor always appears in the combination 
$\Theta^a_I V^I = \kappa^{-1}\,\text{e}^{{\cal K}/2} \, \Theta^a_I U^I$.
The notation $\nabla_\alpha$ stands for K\"ahler-covariant derivatives. Since K\"ahler transformations act as
\begin{equation}
{\cal K}\rightarrow{\cal K} + \lambda(z) + \ov \lambda(\ov z), \qquad
y\rightarrow \text{e}^{\lambda(z)}\, y, \qquad
Z^I (z) \rightarrow \text{e}^{- \lambda(z)}\, Z^I(z),
\end{equation}
the covariant derivatives are
\begin{equation}
\nabla_\alpha \, y = (\partial_\alpha  - {\cal K}_\alpha)y= 0, \qquad\qquad 
\nabla_\alpha U^I = (\partial_\alpha + {\cal K}_\alpha)U^I,
\qquad\qquad
{\cal K}_\alpha = {\partial\over\partial z^\alpha}{\cal K}.
\end{equation}
Supersymmetry imposes the identity
\begin{equation}
\delta^i{}_j \mathscr{V}= \kappa^2 \Bigl(
-3\, S^{ik}\, \ov S_{jk}+W_\alpha{}^{ik}g^{\alpha\ov\beta}\,\ov W_{\ov\beta jk}\Bigr)
+\frac{4}{\kappa^2}N^i_A\ov N_j{}^A ,
\end{equation}
and the gauging and fermion shifts lead then to the following ${\cal N}=2$ scalar potential \cite{ABCDFF, ABCDFFM,FVP}:
\begin{equation}
\label{potential0}
e^{-1}\mathscr{V}=-\frac{1}{2}\left(\text{Im}{\cal N}\right)^{-1IJ}\Theta_I{}^a \Theta_J{}^b P_a^x P_b^x
+ \ov V^I V^J\Theta_I{}^a \Theta_J{}^b
\left(-4\,\kappa^2P_a^xP_b^x+ {2\over\kappa^2}\, h_{uv}\,\xi^u_a\xi^v_b\right)\,,
\end{equation}
where
\begin{equation}
\label{symplectic.identities}
-\frac{1}{2\kappa^2}\left(\text{Im}{\cal N}\right)^{-1IJ}=
\ov V^I V^J+g^{\alpha\bar\beta}\nabla_\alpha V^I\ov\nabla_{\bar\beta} \ov V^J\,,\quad\qquad
\nabla_\alpha V^I=\left(\partial_\alpha+ {\cal K}_\alpha\right)V^I
\end{equation}
in terms of the K\"ahler potential ${\cal K}$.
For later use, we find useful to express the scalar potential \eqref{potential0} in terms of the anti-selfdual covariant 
derivatives $k_{auv}^-$ defined in App.~\ref{appendixquaternionic}:
\begin{equation}
e^{-1}\mathscr{V}=-\frac{1}{2\kappa^4}\left(\text{Im}{\cal N}\right)^{-1IJ}\,\Theta_I{}^a \Theta_J{}^b \, k^-_{auv}k^-_b{}^{uv}
+\frac{\ov V^I V^J\Theta_I{}^a \Theta_J{}^b}{\kappa^2}\left(-4\,k^-_{auv}k^-_{b}{}^{uv}
+ 2\,h_{uv}\,\xi^u_a\xi^v_b\right),
\end{equation}
using the identity 
\begin{equation}
\label{prepotentialsidentity}
P^x_a P^x_b=\frac{1}{\kappa^4}\,k^-_{auv}\,k^-_{b}{}^{uv}\,,
\end{equation}
which can be proved using Eqs.~\eqref{complexJ}.

Supersymmetry breaking is then easily discussed. Firstly, at the ground state defined by the scalar potential, 
nonzero shifts of the spin-$\frac{1}{2}$ fermions indicate the presence of zero, one or two goldstinos, for a spontaneous breaking 
into ${\cal N}=2$, 1 or 0 unbroken supersymmetry(ies). Secondly, if one or two 
supersymmetries remain unbroken, the value of the gravitino shift $S^{ij}$ 
indicates the spacetime geometry of the ground state (AdS or Minkowski).
Partial breaking $ {\cal N}=2\rightarrow {\cal N}=1$  implies that there should be one (and only one) spinor 
$\epsilon_{1i}$ for which {\it three} conditions must be fulfilled:
\begin{equation}
\langle W_z{}^{ij}\rangle\epsilon_{1i}=0, \qquad\qquad
\langle N^i{}_A\rangle\epsilon_{1i}=0, \qquad\qquad
\langle S^{ij}\rangle\epsilon_{1i}= {\mu\over\kappa^2} \, \epsilon_1^i,
\end{equation}
and the scalar curvature of \text{AdS} spacetime is given by $\mathscr{R}=4\Lambda$, $\Lambda=-3|\mu|^2$. 
Furthermore, the second supersymmetry with spinor parameter $\epsilon_2$ should verify either
\begin{equation}
\langle W_z{}^{ij}\rangle\epsilon_{2i} \neq 0 \qquad\text{or}\qquad
\langle N^i{}_A\rangle\epsilon_{2i} \neq 0.
\end{equation}
In the next sections, we analyze these conditions on a special 
K\"ahler geometry with arbitrary prepotential $F(X^I)$ and a generic quaternion-K\"ahler 
geometry for a single hypermultiplet.

\section{The hypermultiplet with isometries and partial breaking}
\label{CPmetric}

For one hypermultiplet, the four-dimensional quaternion-K\"ahler geometry is defined as an Einstein space 
with constant Ricci curvature proportional to $\kappa^2$ \cite{BW}\footnote{For metric $g_{uv}=\kappa^{-2}h_{uv}$ as defined in Eq.~(\ref{kinetichyper}).} and (anti-) selfdual Weyl curvature.
With one or two isometries, metrics for generic quaternion-K\"ahler spaces have been thoroughly discussed. We will use two canonical forms: the Przanowski--Tod (PT) \cite{P, Tod} and the already quoted Calderbank--Pedersen (CP) \cite{CP}.
Both are defined in terms of a solution of a differential equation, nonlinear (Toda) for the PT metric
for spaces with one isometry, linear in the CP metric with two commuting isometries. Since we are interested in the 
latter case, we first consider the hypermultiplet metric in CP coordinates.

\subsection{The CP metric}
\label{CPmetricformulas}

According to Calderbank and Pedersen \cite{CP}, a four-dimensional quaternion-K\"ahler metric with
two commuting isometries can be written in a set of coordinates $\rho>0$, $\eta$, $\psi$ and $\varphi$ 
for every solution $F(\rho,\eta)$ of the linear equation
\begin{equation}
{\partial^2F\over\partial\rho^2}+{\partial^2F\over\partial\eta^2} = {3F\over4\rho^2},
\end{equation}
with isometries acting as shifts of $\psi$ and $\varphi$. The line element 
${\rm d}s^2 = h_{uv} \, {\rm d}q^u {\rm d}q^v$ is
\begin{equation}
\label{CPMetric}
\begin{split}
\text{d}s ^2&=\frac{4\rho^2\left(F_\rho^2+F_\eta^2\right)-F^2}{4F^2}\,\text{d}\ell^2\,
+\frac{\left[(F-2\rho F_\rho)\alpha-2\rho F_\eta\beta\right]^2+\left[(F+2\rho F_\rho)\beta-2\rho F_\eta\alpha\right]^2}
{F^2\left(4\rho^2\left(F_\rho^2+F_\eta^2\right)-F^2\right)}\,,
\end{split}
\end{equation}
where 
\begin{equation}
\alpha=\sqrt{\rho}\text{d}\varphi\,, \qquad \beta=\frac{\text{d}\psi+\eta\text{d}\varphi}{\sqrt{\rho}}, \qquad
\text{d}\ell^2=\frac{\text{d}\rho^2+\text{d}\eta^2}{\rho^2}.
\end{equation}
The metric determinant is
\begin{equation}
{\left(4\rho^2\left(F_\rho^2 + F_\eta^2\right) - F^2\right)^2 \over 16\,\rho^4F^8}
\end{equation}
and positivity requires $4\rho^2\left(F_\rho^2+F_\eta^2\right) > F^2 > 0$.
The CP metric describes  a conformally anti-selfdual Einstein space\footnote{An Einstein metric with 
anti-selfdual Weyl curvature.} with scalar curvature normalized 
to $R=-12$. It is endowed with a triplet of $SU(2)$ selfdual
2-forms $J^x$ (complex structures) which are covariantly constant with an $SU(2)$ connection $\omega^x$ \cite{CP}:
\begin{equation}
\begin{array}{rcl}
J \,\,=\,\, J^x\,i\sigma^x &=& \displaystyle
 \frac{i}{F^2}\left(\left(\rho^2(F_\rho^2+F_\eta^2)-\frac14 F^2\right)\frac{\text{d}\rho\wedge\text{d}\eta}{\rho^2}+\alpha\wedge\beta\right)\sigma_1
\crbig
&&\displaystyle
+\frac{i}{F^2}\left(\left(\rho F_\rho-i\sigma_1\rho F_\eta\right)(\alpha+i\sigma_1\beta)-\frac{1}{2}F(\alpha-i\sigma_1\beta)\right)
\wedge\frac{\text{d}\rho+i\sigma_1\text{d}\eta}{\rho}\sigma_2\,,\crbig
\omega \,\,=\,\, \omega^x \, i\sigma^x &=& \displaystyle
\frac{i}{F}\left(F_\eta\text{d}\rho-\left(\frac12\,F+\rho F_\rho\right)\frac{\text{d}\eta}{\rho}\right)\sigma_1
+\frac{i}{F}\left(\alpha+i\sigma_1\beta\right)\sigma_2\,,
\end{array}
\end{equation}
and the identities \eqref{covJquat}, \eqref{complexJ} are satisfied.

On the metric (\ref{CPMetric}), Calderbank and Pedersen  \cite{CP}
write:\,\footnote{It is the point (\romannumeral2) of their main theorem 1.1.} ``\emph{Any selfdual Einstein metric of nonzero 
scalar curvature with two linearly independent
commuting Killing fields arises locally in this way (\emph{i.e.}, in a neighbourhood of any point, it is of the form (3.2) 
up to a constant multiple)}.'' We will see that a slight inaccuracy in this statement allows for an exception 
which is of fundamental importance in our subject. 

The two Killing vectors of the CP metric are by construction $\xi_1=\partial_\varphi$ and $\xi_2=\partial_\psi$.
Using Eq.~\eqref{Killing.prepotential}, the triplets of Killing prepotentials (or moment maps) are
\begin{equation} 
P^x_1=\frac{1}{\kappa^2\sqrt{\rho}F}\begin{pmatrix}
   0 \\
   -\,\rho \\
   \eta
  \end{pmatrix} , \qquad
P^x_2=\frac{1}{\kappa^2\sqrt{\rho}F}\begin{pmatrix}
  0 \\
   0\\
   1
  \end{pmatrix}.
\end{equation}
The standard vierbein one-forms of the metric \eqref{CPMetric} are
\begin{equation}
\begin{array}{rclrcl}
e^0 &=&\displaystyle \frac{\sqrt{4\rho^2(F_\rho^2+F_\eta^2)-F^2}}{2F}\,\frac{\text{d}\rho}{\rho}\,,\quad&\quad
e^1 &=&\displaystyle \frac{\sqrt{4\rho^2(F_\rho^2+F_\eta^2)-F^2}}{2F}\,\frac{\text{d}\eta}{\rho}\,,\crbig
e^2 &=&\displaystyle \frac{(F-2\rho F_\rho)\alpha-2\rho F_\eta\beta}{F\sqrt{4\rho^2(F_\rho^2+F_\eta^2)-F^2}}\,,
 &
e^3 &=&\displaystyle \frac{(F+2\rho F_\rho)\beta-2\rho F_\eta\alpha}{F\sqrt{4\rho^2(F_\rho^2+F_\eta^2)-F^2}}\,.
\end{array}
\end{equation}
We will need the corresponding symplectic vielbeins $f^{iA}{}_u$ obtained from relations
\begin{equation}
\text{d}s^2 = \delta_{mn} \, e^m e^n = \varepsilon_{ij} \Omega_{AB} \,
f^{iA}{}_uf^{jB}{}_v \, \text{d}q^u \text{d}q^v
\end{equation}
or
\begin{equation}
f^{iA}{}_u =\frac{1}{\sqrt{2}}\left(e^0_u\,\varepsilon\pm i\,
e^x_u\,\varepsilon\,\sigma^x\right)^{iA}\,,\qquad x=1,2,3\,,
\end{equation}
and we have checked that the $f^{iA}{}_u$'s satisfy Eq.~\eqref{metric.symplectic}. We will use the $+$ sign below.

\subsection{Fermion shifts}\label{secshift}

Consider a generic dyonic gauging of the two isometries, described by:
\begin{equation}
\label{KillingsSOgen}
\Theta_I{}^a= \begin{pmatrix}
   g_0 & g_1 \\
   0 & g_2 \\
 0 & 0 \\
 0 & -g_3 
  \end{pmatrix} \,, \qquad\quad a=1,2\,,
\end{equation}
where the embedding tensor $\Theta_I{}^a$ is compatible with the locality condition \eqref{embeddingconsistency} and
$g_{0,1,2,3}$ are the gauge couplings.
We use a prepotential frame and formulas (\ref{Fframe}) apply. The corresponding K\"ahler potential is
given in Eq.~(\ref{Kis}) for a single $z$:
\begin{equation}
\label{genericlocalKahler}
{\cal K} = - \ln{\cal Y}, \qquad\qquad {\cal Y}= 2(f+\bar f)-(z-\bar z)(f_z-\ov f_{\ov z}) \,.
\end{equation}
In order to evaluate the fermion shifts given in \eqref{fermion}, we use the results of \S~
\ref{CPmetricformulas}. Defining
\begin{equation}
\label{ctilde}
\widetilde c=g_1+g_2 z+i g_3 f_z+g_0\eta,
\end{equation}
we find
\begin{eqnarray}
\nonumber
S^{ij} &=& \displaystyle - {1\over\kappa} \text{e}^{{\cal K}/2}
\left(g_0P_1^2\delta^{ij}+i\left(\left(g_1+g_2 z+i g_3 f_z \right)P_2^3+g_0P_1^3\right)(\sigma_1)^{ij}\right)\\
&=& \displaystyle
-\frac{\text{e}^{{\cal K}/2}}{\kappa^3\sqrt{\rho}F}
\left(-g_0\rho\,\delta^{ij}+i\widetilde c(\sigma_1)^{ij}\right)
\end{eqnarray}
for the gravitino shift, and
\begin{eqnarray} 
\nonumber
&\displaystyle W_z^{ij} = - {1\over\kappa}\,\text{e}^{{\cal K}/2}\,\Theta_I{}^a P_a^{ij}\nabla_z U^I=
i\,\text{e}^{{\cal K}/2}\frac{(g_2+i g_3 f_{zz})}{\kappa^3\sqrt{\rho}F}(\sigma_1)^{ij} - {\cal K}_z\, S^{ij}\,,&
\\
&\displaystyle N^i{}_A=-\frac{\text{e}^{{\cal K}/2}}{\kappa\sqrt{2\rho}F\sqrt{4\rho^2(F_\rho^2+F_\eta^2)-F^2}}
\left(\rho A_2\sigma_2+A_3\sigma_3\right)^{iA}\,,&\nonumber
\\
&\quad A_2=-g_0F+2(\widetilde c F_\eta+g_0\rho F_\rho)\,,\quad\qquad
A_3=\widetilde c F+2\rho(-g_0\rho F_\eta+\widetilde c F_\rho)&
\label{CPhyperino}
\end{eqnarray}
for the shifts of spin-$\frac{1}{2}$ fermions.\footnote{The fermion shifts have dimension mass$^3$ ($S$ and $W$)
or mass$^1$ ($N)$.}

\subsection{Partial breaking in flat space}\label{secpart}

The first condition for partial breaking is certainly that the ground state does not lead to two goldstinos. The
determinants of $W_z^{ij}$ and $N^i{}_A$ should vanish at the ground state.
Cancelling the determinant of $W_z^{ij}$ requires
\begin{equation}
\label{det0}
ig_2 - g_3\langle f_{zz}\rangle = \mp \langle ( \pm i  \widetilde c + g_0 \, \rho ){\cal K}_z\rangle 
\end{equation}
with zero eigenvector\,\footnote{The $\pm$ or $\mp$ signs are correlated between the various 
equations. The parameter function $v(x)$ has dimension mass$^{-1/2}$.}
\begin{equation}
\widehat\epsilon_i (x) = \begin{pmatrix}
\mp 1 \\ 1 \end{pmatrix} v(x),
\end{equation}
and this $\widehat\epsilon_i$ is also eigenvector of $S^{ij}$ with
\begin{equation}
\langle S^{ij}\rangle \, \widehat\epsilon_j = \left< \frac{\text{e}^{K/2}}{\kappa^3\sqrt{\rho}F} \right>
\langle g_0\rho \pm i\widetilde c \rangle
\widehat\epsilon_i.
\end{equation}
The eigenvalue should vanish for a Minkowski ground state:
\begin{equation}
\label{Mink}
g_0 \langle\rho\rangle = \mp i \langle\widetilde c \rangle  \ne 0 \qquad\qquad \makebox{(Minkowski)}
\end{equation}
turning condition (\ref{det0}) into
\begin{equation}
g_2 + ig_3\langle f_{zz}\rangle = 0 \qquad\qquad \makebox{(Minkowski)},
\end{equation}
and then to 
\begin{equation}
\left\langle W_z^{ij} \right\rangle =  - \left\langle {\cal K}_z S^{ij} \right\rangle
\qquad\qquad \makebox{(Minkowski)}.
\end{equation}
The determinant of $N^i{}_A$ turns out to be  proportional to 
\begin{equation}
\left(\rho^2 A_2\right)^2 + A_3^2 = 8 \, \widetilde c^2 \rho F (F_\rho \pm i F_\eta),
\end{equation}
using the Minkowski conditions (\ref{Mink}) and (\ref{det0}). The conditions for the positivity of the 
CP metric, $\rho, F, F^2_\rho + F^2_\eta > 0$ and condition (\ref{Mink}) imply that $N^i{}_A$
does not have a zero eigenvalue.
Hence, t\emph{he partial breaking of ${\cal N}=2$ supersymmetry in Minkowski 
spacetime is excluded whenever the hypermultiplet can be described in the
CP field coordinates and metric.} According to Ref.~\cite{CP}, this would be always the case.

There is an apparent contradiction between this conclusion and the known existence \cite{FGP}
of a partial breaking on the $SO(4,1)/SO(4)$ hypermultiplet in Minkowski spacetime.  We will see shortly that
for this quaternion-K\"ahler space, and only for this space, there exists a pair of isometries for which the 
coordinates used by Calderbank and Pedersen \cite{CP} \emph{do not} exist. This is the earlier quoted exception, leading to a statement of \emph{uniqueness} 
for partial breaking with a single multiplet and two gauged isometries.

\subsection{Partial breaking in AdS}\label{secpartADS}

The obstruction found for partial breaking into Minkowski spacetime does not exist for 
AdS ground states. Partial breaking in this case requires at the first place Eq.~(\ref{det0}). With this condition, 
the gaugino and gravitino shifts read
\begin{equation} 
\begin{array}{rcl}
\langle W_z^{ij} \rangle &=&\displaystyle - \left< {g_0\sqrt\rho \text{e}^{{\cal K}/2} \over\kappa^3 F}\,{\cal K}_z \right> \,
\begin{pmatrix}
1 & \pm1 \\ \pm1 & 1
\end{pmatrix},
\crbig
\langle S^{ij} \rangle 
&=&\displaystyle \left<  {\text{e}^{{\cal K}/2} \over\kappa^3 \sqrt\rho F} \right> \left[  \mp i\langle\widetilde c\rangle
\begin{pmatrix}
1 & \pm1 \\ \pm1 & 1
\end{pmatrix} 
+ \langle g_0\rho\pm i\widetilde c \rangle 
\begin{pmatrix}
1 & 0 \\ 0 & 1
\end{pmatrix}
\right]\end{array}
\end{equation}
at the ground state.
For the unbroken supersymmetry parameter $\widehat\epsilon$ (the zero eigenvector of $\langle W^{ij}
\rangle$),
\begin{equation}
\delta\, \psi_\mu^i = \left<  {\text{e}^{{\cal K}/2} \over 2\kappa \sqrt\rho F} (g_0\rho\pm i\widetilde c) \right>\, 
\gamma_\mu \widehat\epsilon^i + \cdots,
\end{equation}
and the cosmological constant is
\begin{equation}
\Lambda = -3\, \left< {\text{e}^{\cal K}\over\kappa^2\rho F^2} \left(g_0^2 \, \rho^2 + |\widetilde c|^2\right)\right> .
\end{equation}
The Minkowski condition (\ref{Mink}) which cancels $\Lambda$ does not apply and
the second condition for partial breaking is that $\widehat\epsilon$ is also a zero eigenvector 
of the hyperino shift matrix (\ref{CPhyperino}):
\begin{equation}
\label{hyper2}
\left< \rho A_2 \rangle = \mp i \langle A_3 \right>.
\end{equation}
Solutions to Eqs.~(\ref{det0}) and (\ref{hyper2}) would lead to stable AdS ground 
states.\,\footnote{Their stability is carefully discussed in the appendix B of Ref.~\cite{LST1}.
{  An early example was given in ref.~\cite{CGP4}.}}

\paragraph{An example.}

We can realize the above conditions for ${\cal N}=1$ AdS vacua in a specific example. We consider for this a CP metric with
$F_\eta=0$, {\it i.e.}
\begin{equation}
F=\frac12\,\rho^{3/2}-\sigma\rho^{-1/2}\,,\qquad \qquad \sigma=\text{constant}\,.
\end{equation}
This metric has extended isometry $\text{Heisenberg} \ltimes U(1)$ and it describes the scalar manifold of 
the universal hypermultiplet in type II strings, including the one-loop perturbative corrections, as
obtained in Ref.~\cite{AMTVoneloop}. The case $\sigma=0$ is the tree-level $SU(2,1)/SU(2)\times U(1)$.

From the expressions of $A_2$ and $A_3$ in Eq.~(\ref{CPhyperino}), one obtains:
\begin{eqnarray}
A_2 &=& g_0(-F+2\rho F_\rho)=g_0\left(\rho^{3/2}+2\,\sigma\rho^{-1/2}\right),
\\
A_3 &=& \widetilde{c}(F+2\rho F_\rho) = 2\,\widetilde{c}\rho^{3/2},
\end{eqnarray}
and thus the condition (\ref{hyper2}) implies:
\begin{equation}
\label{ctildesol}
{\rm Re}\,\widetilde{c}=0\,, \qquad\qquad {\rm Im}\,\widetilde{c}=\pm g_0\left({\rho\over 2}+{\sigma\over\rho}\right),
\end{equation}
where we dropped the symbols of expectation values. On the other hand, condition (\ref{det0}) yields
\begin{equation}
-g_2+g_3\, {\rm Im}\,f_{zz} =\pm g_0\rho\,{\rm Im}\, {\cal K}_z \,, \qquad\qquad 
{\rm Re}\,\widetilde{c}=-g_3\, {\rm Re}\, f_{zz} \pm g_0\rho \,{\rm Re}\, {\cal K}_z\,.
\end{equation}
It follows that there are four equations that can be solved for the four expectation values of $\rho$, $\eta$ and $z$.
Indeed, using the expression of $\widetilde{c}$ in Eq.~(\ref{ctilde}), one obtains:
\begin{eqnarray}
\label{etarho}
g_0\eta &=& -g_1-g_2{\rm Re}\, z+g_3\,{\rm Re}\, f_z, \\
\pm g_0\rho\, {\rm Im}\,{\cal K}_z &=& g_2-g_3\, {\rm Im}\, f_{zz},
\end{eqnarray}
which can be used in the remaining two equations (right part of (\ref{ctildesol}) and (\ref{etarho})) for determining $z$.
For instance, for $g_3=0$, on finds the solution:
\begin{equation}
\begin{array}{rclrcl}
g_0\eta &=& -g_1-g_2 \,{\rm Re}\, z\,,  \quad&\quad \pm g_0\rho &=& 2 g_2 \, {\rm Im}\, z\,,
\crbig
{\rm Re}\, {\cal K}_z &=& \displaystyle{1\over 2}\,, \quad&\quad 
{\rm Im}\,{\cal K}_z &=& \displaystyle -{1\over 2\,{\rm Im} z}\,,
\end{array}
\end{equation}
where the last two equations determine $z$, using the expression (\ref{genericlocalKahler}) for ${\cal K}$. 
Note that partial ${\cal N}=2$ supersymmetry breaking in AdS can be realised without introducing magnetic 
FI coupling terms $(g_3=0)$.

\subsection{Hyperbolic space and Calderbank--Pedersen coordinates}
\label{H2CP}

We now come back to the issue met at the end of \S~\ref{secpart}.
To resolve the case, we will show that the hyperbolic space cannot be described by a CP metric with 
shift isometries on $(\varphi,\psi)$ generated by pairs of elements in the three-dimensional abelian subalgebra of 
$SO(4,1)$.\footnote{See App.~\ref{appendixhyperbolic} for a review on $\text{H}_4$ coordinates.}
In other words, CP coordinates do not exist for this case.
To proceed, we simply compare the value of scalar quantities (independent on the choice of coordinates), 
calculated either in CP or in PT coordinates.

The Calderbank--Pedersen metric (\ref{CPMetric}) has a well-defined pair of isometries and Killing vectors. 
Scalar quantities like those appearing in the identity \eqref{prepotentialsidentity} can then be calculated unambiguously,
\begin{equation}
\label{QKtoCP}
k^-_{1uv}\,k^-_{1}{}^{uv}=\frac{\rho^2+\eta^2}{\rho\,F^2}\,,\qquad
k^-_{1uv}\,k^-_{2}{}^{uv}=\frac{\eta}{\rho\,F^2}\,,\qquad
k^-_{2uv}\,k^-_{2}{}^{uv}=\frac{1}{\rho\,F^2},
\end{equation}
and the dependence on the quaternion-K\"ahler space is in the function $F(\rho,\eta)$ only. 

Consider now the simplest quaternion-K\"ahler space, 
\begin{equation}
{Sp(2,2) \over Sp(2)\times Sp(2)} \sim \frac{SO(4,1)}{SO(4)}.
\end{equation}
This hyperbolic space admits coordinates in which the line element is 
\begin{equation}
\label{metricSO}
\text{d}s^2= {1\over b_0^2}\left(\text{d}b_0^2+\text{d}b_1^2+\text{d}b_2^2+\text{d}b_3^2\right)\,.
\end{equation}
This is a conformally-flat space with $R_{uv}=-3\, h_{uv}$.
The corresponding symplectic vierbeins
\begin{equation}
f^{iA}_0=-\frac{1}{\sqrt{2}b_0}\tau_1^{iA},\qquad
f^{iA}_1=-\frac{1}{\sqrt{2}b_0}\tau_2^{iA},\qquad
f^{iA}_2=-\frac{1}{\sqrt{2}b_0}\tau_3^{iA},\qquad
f^{iA}_3=\frac{1}{\sqrt{2}b_0}\varepsilon^{iA},
\end{equation} 
where $(\tau_x)^{iA} = (i\varepsilon\sigma_x)^{iA}$,
follow from their definition \eqref{metric.symplectic}.
The triplet of $SU(2)$ self-dual two-forms $J^x$ (complex structures)
\begin{equation}
\begin{split}
&J^1=\frac{1}{b_0^2}\left(\text{d}b_0\wedge\text{d}b_3 + \text{d}b_1\wedge\text{d}b_2\right)\,,\\
&J^2=\frac{1}{b_0^2}\left(\text{d}b_0\wedge\text{d}b_2 + \text{d}b_3\wedge\text{d}b_1\right)\,,\\
&J^3=-\frac{1}{b_0^2}\left(\text{d}b_0\wedge\text{d}b_1 + \text{d}b_2\wedge\text{d}b_3\right)
\end{split}
\end{equation}
is covariantly constant up to the $SU(2)$ connection $\omega^x$
\begin{equation}
\omega^1=-\frac{\text{d}b_3}{b_0}\,,\qquad \omega^2=-\frac{\text{d}b_2}{b_0}\,,\qquad 
\omega^3=\frac{\text{d}b_1}{b_0}\,.
\end{equation}
Conditions \eqref{covJquat} and \eqref{complexJ} are verified.
We are interested in the Killing vectors of two translation isometries acting on $b_2$ and $b_3$:
\begin{equation}
\xi_1=\partial_{b_2}\,, \qquad\qquad \xi_2=\partial_{b_3}\,.
\end{equation}
Their Killing prepotential triplets follow from Eq.~\eqref{Killing.prepotential}:
\begin{equation}
P_1=-\frac{1}{\kappa^2b_0}\begin{pmatrix}
0 \\ 1 \\ 0 \end{pmatrix},\qquad\qquad
P_2=-\frac{1}{\kappa^2b_0}\begin{pmatrix} 1 \\ 0 \\ 0 \end{pmatrix}.
\end{equation}
In these coordinates and for these isometries, we find the following expression for the scalars
appearing in Eq.~(\ref{prepotentialsidentity}):
\begin{equation}
k^-_{1uv}\,k^-_{1}{}^{uv}=\frac{1}{b_0^2},\qquad\qquad
k^-_{1uv}\,k^-_{2}{}^{uv}=0,\qquad\qquad
k^-_{2uv}\,k^-_{2}{}^{uv}=\frac{1}{b_0^2}.
\end{equation}
The comparison with the generic values (\ref{QKtoCP}) obtained for the CP metric indicates that
for these isometries, CP coordinates cannot be found.\,\footnote{A similar conclusion, technically more involved though, 
derives from comparing the inner products of the two isometries, {\it i.e.} $\xi^u_a\xi^v_b h_{uv}$.}

The origin of this obstruction is located in the derivation of the CP metric given in Ref.~\cite{CP}.
This metric is a consequence of the Joyce description for anti-selfdual conformal metrics with 
a $U(1)\times U(1)$ symmetry \cite{Joyce}, 
the Jones--Tod correspondence for four dimensional anti-selfdual spaces with at least one isometry \cite{JonesTod}, and the use of Przanowski--Tod (PT) theorem to determine which metrics, among the 
Joyce metrics, are Einstein spaces \cite{P, Tod, Tod2}.
In short, to identify the Einstein representatives among the conformal structures with anti-selfdual Weyl tensor, 
one employs the PT form where the metric is generated by a function $\Psi(X,Y,Z)$ solving the continual Toda equation  \cite{P,Tod,Tod2}\footnote{As usual, indices indicate derivatives.}
\begin{equation}
\Psi_{XX}+\Psi_{YY}+ (\text{e}^\Psi)_{ZZ}=0.
\end{equation}
The PT line element is then
\begin{equation}
\begin{split}
\label{prztod1}
&\text{d}s^2=\frac{1}{Z^2}\left[\frac{1}{U}(\text{d}\psi+\omega)^2+
U\left(\text{d}Z^2+\text{e}^\Psi\left(\text{d}X^2+\text{d}Y^2\right)\right)\right],
\crbig
&\text{d}\omega= U_X \, \text{d}Y\wedge \text{d}Z+ U_Y \, \text{d}Z\wedge \text{d}X+
(U\, \text{e}^\Psi)_Z\,\text{d}X\wedge \text{d}Y\,,\crbig
&2 U=2-Z\, \Psi_Z\,.
\end{split}
\end{equation}
The quaternion-K\"ahler metric has one isometry acting on the fourth coordinate $\psi$ and generated by 
$\partial_\psi$. The simplest solution is of course
\begin{equation}
\Psi = C= \makebox{constant}, \qquad U=1, \qquad {\rm d}\omega=0, \qquad \omega={\rm d}g(X,Y,Z),
\end{equation}
for which 
\begin{equation}
\text{d}s^2=\frac{1}{Z^2}\left[ \text{d}(\psi+g)^2+
\text{d}Z^2+\text{e}^C\left(\text{d}X^2+\text{d}Y^2 \right)\right].
\end{equation}
This is the metric (\ref{metricSO}) with $b_0=Z$, $b_1= \text{e}^{C/2}X$, $b_2=\text{e}^{C/2}Y$, $b_3=\psi+g$ and
the isometry of the PT metric shifts $b_3$.

In order to make contact with the CP metric, we assume the existence, in the PT description,
of a second isometry generated 
by $\partial_\varphi$. The case of primary interest for us is a shift isometry acting on $b_1$ or $b_2$. 
We choose a translation isometry of the $Y$ coordinate.

Assume then that $\varphi=Y$ and that $\Psi$ does not depend on $Y$. Finding the CP coordinates is possible using a transformation due to Ward \cite{Ward}:\footnote{See App.~\ref{secWard} for details.} 
\begin{equation}
\begin{split}
&\left(X, \, Z; \, \text{e}^{\Psi(X,Z)}\right) \qquad\Longrightarrow\qquad (\rho,\, \eta; \, V(\rho,\eta))\,,\crbig
&X=V_\eta\,,\qquad 2\,Z=\rho\,V_\rho\,,\qquad {1\over4}\,\rho^2=\text{e}^\Psi\,,\crbig
&\Psi_{XX}+\left(\text{e}^\Psi\right)_{ZZ}=0\qquad\Longrightarrow\qquad \frac{1}{\rho}\left(\rho V_\rho\right)_\rho
+ V_{\eta\eta}=0\,,
\end{split}
\end{equation}
resulting in the CP metric \eqref{CPMetric}, with $F=\sqrt{\rho}\, V_\rho$. This transformation is clearly 
incompatible with a constant $\Psi$. For this hypermultiplet manifold and for this choice of second isometry with
Killing vector $\partial_Y$, CP coordinates $\rho$ and $\eta$ do not exist and the argument against partial
breaking proved in the previous section 
does not hold.
The constancy of the Toda potential is at the origin of this exception.

With $SO(4,1)$ isometry, the hyperbolic space has a variety of other inequivalent pairs of commuting isometries.
For these pairs, the corresponding Toda potentials are not constant and CP coordinates do exist. Some examples of CP 
coordinates for other isometries of the hyperbolic space are described in App.~\ref{appendixhyperbolic}.

\newcommand{\skipthispart}[1]{}
\skipthispart{
\item
Let us now consider $\varphi=\vartheta$, where $\vartheta$ is the polar angle in the $(X,Y)$ plane, and 
we apply Ward's transformation at \eqref{prztod1} 
\begin{equation}
\begin{split}
&(r,Z;\text{e}^{\Psi(r,Z)}) \quad\mapsto\quad (\rho,\eta;\Phi(\rho,\eta))\,,\\
&\ln r=\Phi_\eta\,,\quad Z=\rho\,\Phi_\rho\,,\quad \rho^2=r^2\text{e}^\Psi\,,\\
&\frac{1}{r}\left(r\,\Psi_r\right)_r+\left(\text{e}^\Psi\right)_{ZZ}=0\quad\Longrightarrow
\quad \frac{1}{\rho}\left(\rho\,\Phi_\rho\right)_\rho+\Phi_{\eta\eta}=0\,,
\end{split}
\end{equation}
resulting again to the CP metric \eqref{CPMetric}, with $F=\sqrt{\rho}\,\Phi_\rho$.

Note that the above transformation is non singular, even for constant $\Psi$,
and \eqref{prztod1} truncates to the hyperbolic space
\begin{equation}
\text{d}s^2_{\text{QK}}=\frac{1}{Z^2}\left(\text{d}\psi^2+
\text{d}Z^2+\text{d}r^2+r^2\text{d}\varphi^2\right)\,,
\end{equation}
whose corresponding CP form \eqref{CPMetric} reads
\begin{equation}
\label{constantPsipolar}
\rho=c_1 r\,,\quad \eta=Z-c_2\,,\quad F=\frac{\eta+c_2}{\sqrt{\rho}}\,,
\end{equation}
where $\text{e}^\Psi=c_1^2$ and $c_{1,2}$ are constants.

}

\section{Partial breaking and the APT model}
\label{SO14Partial}

\boldmath
\subsection{The $SO(4,1)/SO(4)$ model}\label{secSO4}
\unboldmath
Ferrara, Girardello and Porrati (FGP) \cite{FGP, FGP2} have shown that partial breaking occurs on the simplest 
quaternion-K\"ahler space for one hypermultiplet, $SO(4,1)/SO(4)$, with two gauged translation isometries.
Explicitly, coordinates  (\ref{metricSO}) with
\begin{equation}
\text{d}s^2={1\over b_0^2}\left(\text{d}b_0^2+\text{d}b_1^2+\text{d}b_2^2+\text{d}b_3^2\right), \qquad\qquad
{\cal L}_{\text{\text{kin.}} } = -{e\over2(\kappa b_0)^2} (\partial_\mu b^u) (\partial^\mu b^u),
\end{equation}
and Killing vectors
\begin{equation}
\xi_1=\partial_{b_2}\,,  \qquad\qquad \xi_2=\partial_{b_3}
\end{equation}
are used for constructing the ${\cal N}=2$ supergravity lagrangian. In Ref.~\cite{FGP} they first worked in the 
{\it non-prepotential} frame described in Sec.~\ref{preliminaries}, Eqs.~(\ref{exc}) and (\ref{exc2}). Then
they reworked the example in a generic 
frame with arbitrary prepotential function $f(z)$ \cite{FGP2}.

Our objective in this section is to complete the description of the model by showing explicitly that the 
${\cal N}=2$ supergravity theory (at finite $\kappa$ then) admits a stable
ground state with partial breaking which continuously deforms to the APT model in the gravity-decoupling limit
$\kappa\rightarrow0$. This can be seen as deriving {\it off-shell} the APT lagrangian as the $\kappa\rightarrow0$ 
limit of the $SO(4,1)/SO(4)$ supergravity lagrangian.\footnote{Following for instance Ref.~\cite{Aetal}.
Although the statement exists in the 
literature, we have not found an explicit
construction with an appropriate use of the concept of prepotential frame.}

Using the embedding tensor (\ref{KillingsSOgen}), the prepotential frame (\ref{Fframe}) leading to K\"ahler 
potential (\ref{Kis}) and coordinates $b^u$ with metric (\ref{metricSO}) for the hypermultiplet, the supergravity 
potential reads:
\begin{equation}
\label{potentialAPTSUGRA}
\mathscr{V}= \frac{\text{e}^{\cal K} }{\kappa^4\,b_0^2 {\cal K}_{z\bar z}}
\Bigl[ \left(g_0^2+|c|^2\right)(-{\cal K}_{z\bar z}+{\cal K}_z {\cal K}_{\bar z})+
|c_z|^2+\bar c c_z {\cal K}_{\bar z}+c\bar c_z {\cal K}_z \Bigr]
\end{equation}
with $c$ defined as
\begin{equation}
c=-i(g_1+g_2 z+i g_3f_z ),
\end{equation}
and $c_z = -ig_2 + g_3 f_{zz}$.
Since hypermultiplet scalars only appear in the prefactor $b_0^{-2}$, the ground state 
of the potential, in order to escape the runaway of $b_0$, requires Minkowski geometry, $\langle \mathscr{V} \rangle=0$.
In Ref.~\cite{FGP}, the authors  consider the particular case $g_1=g_2=0$, $f(z) \sim z$, ${\cal K}=-\ln(z+\ov z)$ and then
$\mathscr{V} \equiv0$. 

Notice that the scalar potential (\ref{potentialAPTSUGRA}) vanishes if
\begin{equation}
\langle c_z \rangle = 0 \qquad\Longrightarrow\qquad g_3 \langle f_{zz} \rangle = ig_2 .
\end{equation}
Since $\langle f_{zz} \rangle$ is imaginary, 
$\langle{\cal K}_{z\ov z}\rangle = \langle{\cal K}_z{\cal K}_{\ov z}\rangle$ and $\langle \mathscr{V} \rangle=0$.


\boldmath
\subsection{${\cal N}=1$ Minkowski vacua}\label{secSO4Mink}
\unboldmath

The fermion shifts (\ref{fermion}) induced by this gauging read
\begin{equation}
\begin{split}
S^{ij}=\frac{\text{e}^{{\cal K}/2}}{\kappa^3b_0}\,\left(g_0\,\mathbb{I}_2+c\,\sigma_3\right)^{ij}\,, \qquad\qquad
N^i{}_A=\frac{\text{e}^{{\cal K}/2}}{\kappa\sqrt{2} b_0}\left(g_0\,\sigma_3+c\,\mathbb{I}_2\right)^{iA} 
\end{split}
\end{equation}
for gravitinos and hyperinos. They verify the relation\,\footnote{As matrices, $\kappa^2S=\sqrt2 \sigma_3 N$.} 
\begin{equation}
S^{ik}\, \ov S_{jk}=\frac{2}{\kappa^4}N^i_A\ov N_j{}^A\,.
\end{equation}
For gauginos,
\begin{equation}
W_z^{ij} = - \kappa^{-1} \text{e}^{{\cal K}/2}\,\Theta_I{}^a P_a^{ij}\nabla_z U^I
= -\text{e}^{{\cal K}/2}\frac{c_z}{\kappa^3 b_0}(\sigma_3)^{ij}-{\cal K}_z\, S^{ij}\,,
\end{equation}
where we have used
\begin{equation}
\Theta_I{}^a\partial_z U^I= \begin{pmatrix}
0 \\ ic_z \end{pmatrix} .
\end{equation}
The conditions for partial breaking are then easily stated. To have a common zero eigenvector for $W^{ij}$
and $N^i{}_A$, we need a dyonic (electric {\it and} magnetic) gauging with $g_3 \ne 0 \ne g_0$ and
\begin{equation}
\label{parteq1}
\langle g_3 f_z - i g_2 z \rangle = \pm g_0 + ig_1\,, 
\qquad\qquad
g_3 \langle f_{zz} \rangle = ig_2 .
\end{equation}
The first condition $\langle c \rangle = \pm g_0$ leads to a zero eigenvector $\widehat\epsilon$ of $N^i{}_A$ while the second condition $\langle c_z\rangle = 0$ ensures that the same $\widehat\epsilon$ is a 
zero eigenvector of $W_z^{ij}$. This second condition for partial breaking also implies  
$\langle S^{ij}\rangle\widehat\epsilon_j=0$ and $\langle\mathscr{V}\rangle = 0$ and then partial breaking
can only exist in Minkowski spacetime.
The conditions (\ref{parteq1}) define an ${\cal N}=1$ supersymmetric stable ground state. 
In Ref.~\cite{FGP} where $g_1=g_2= f_{zz}=0$, these conditions reduce to $g_3=\pm g_0 \ne 0$. 

Solving the conditions for partial breaking commonly impose, for a given choice of $f(z)$, particular values or 
relations on the coupling constants. For instance, a linear $f=z$, as used in Ref.~\cite{FGP}, has partial 
breaking only if $g_0=\pm g_3\ne0$, $g_1=g_2=0$. The conditions may be impossible to solve: 
$f=z^2$ forces all $g_i$ to be zero (but this example is irrelevant since the K\"ahler metric ${\cal K}_{z\ov z}
\equiv0)$. For a generic prepotential $f(z)$, one usually finds that two couplings are determined in terms of the other 
two.

The spectrum of the partially broken theory includes ${\cal N}=1$ supergravity ($2_\text{B}+2_\text{F}$ 
on-shell states),
a massive ${\cal N}=1$ gravitino multiplet (gravitino, the two spin-one fields, one fermion, $6_\text{B}+6_\text{F}$),
a massless chiral multiplet ($2_\text{B}+2_\text{F}$) and a chiral multiplet with the scalar $z$ and mass proportional to
the free parameter $\langle f_{zzz} \rangle$, precisely as in the APT model, see below. 
The four hypermultiplet scalars are massless (two are Goldstone bosons)
and the mass matrix reduces to $z$ only with
\begin{equation}
\langle \mathscr{V}_{z\bar z}\rangle=
\left\langle \frac{g_3^4{\cal Y}}{4 \kappa ^4g_0^2b_0^2}\, |f_{zzz}|^2 \right\rangle  \geqslant0\, ,
\end{equation}
where ${\cal Y}$ is defined through the K\"ahler potential \eqref{genericlocalKahler}:
\begin{equation}
 {\cal K} =-\ln{\cal Y}\,, \qquad\qquad
{\cal Y}=-\frac{i(z-\bar z)(2g_1+g_2(z+\bar z))}{g_3}+2(f+\bar f)>0\,.
\end{equation}
Hence, the mass of the scalar $z$ (and of its fermion partner) is given by
\begin{equation}
\label{massscalar}
m_z^2 = \kappa^2 \left< { \mathscr{V}_{z\bar z} \over {\cal K}_{z\bar z} } \right> 
= \left< \frac{g_3^6{\cal Y}^3}{16 \, \kappa^2 g_0^4b_0^2}\, |f_{zzz} |^2 \right>
\end{equation}
since ${\cal K}_{z\bar z}= \frac{4g_0^2}{ g_3^2{\cal Y}^2}$.

At the ${\cal N}=1$ ground state, the value of the hypermultiplet scalar $\langle b_0\rangle$ is an 
arbitrary parameter. From the gravitino shift matrix\,\footnote{Which in a Minkowski ground state is 
proportional to the mass matrix, $S^{ij} \sim \kappa^{-2} m_{\frac{3}{2}}^{ij}$.} or from the expression of the 
scalar potential however,
the mass of the massive gravitino scales as
\begin{equation}
m_{\frac{3}{2}} \sim \langle \kappa b_0 \rangle^{-1},
\end{equation}
and the theory has two order parameters, $\langle b_0 \rangle$ and $\langle f_{zzz} \rangle$ 
for the massive gravitino and chiral (with $z$) multiplets respectively.

In order to discuss the gravity-decoupling limit $\kappa\rightarrow0$ of the supergravity theory and make contact with the APT model, we first redefine
the hypermultiplet scalars ($\langle b_0\rangle\ne0$):\,\footnote{This is a simplistic use of the procedure 
described in Refs.~\cite{AADT,ADPS1, ADPS2, ABL}.}
\begin{equation}
\label{rigidSO14}
b_0 =  \langle b_0\rangle (1 + \kappa \tilde\mu \,\widetilde b_0)\,, \qquad\qquad
b_i =   \kappa\tilde\mu \langle b_0\rangle \, \widetilde b_i \,,
\qquad\qquad i=1,2,3,
\end{equation}
where $\tilde\mu$ is a mass scale (and $\kappa\tilde\mu \sim \frac{\tilde\mu}{M_\text{P}}$ is dimensionless). The 
hypermultiplet kinetic terms are then
\begin{equation}
{\cal L}_{\text{hyper}} = -{e\, \tilde\mu^2\over2\left(1 + \kappa\tilde\mu \widetilde b_0\right)^2} \,\delta_{uv} 
(\partial_\mu \widetilde b^u)(\partial^\mu \widetilde b^v),
\end{equation}
and in the limit $\kappa\rightarrow0$, the kinetic metric is the trivial hyper-K\"ahler $h_{uv}=\tilde\mu^2 \delta_{uv}$.\,\footnote{
We could as well define dimension-one fields with $\tilde \mu\widetilde b^u \rightarrow \widetilde b^u$.}

For the vector multiplet kinetic term, we need as $\kappa\rightarrow0$
\begin{equation}
-{e\over\kappa^2} \, {\cal K}_{z\ov z} (\partial_\mu z)(\partial^\mu\ov z) 
\qquad\longrightarrow\qquad
- (i\ov{\cal F}_{\ov{xx}} - i{\cal F}_{xx}) (\partial_\mu x)(\partial^\mu\ov x),
\end{equation}
where ${\cal F}(x)$ is the dimension-two prepotential of the rigid ${\cal N}=2$ theory and $x$ is a dimen\-sion-one scalar. 
In other words, we need 
\begin{equation}
{1\over\kappa^2}\, {\cal K}(z,\ov z) = - {1\over\kappa^2}\, \ln{\cal Y} \qquad\longrightarrow\qquad 
-i\ov x {\cal F}_x + ix\ov{{\cal F}}_{\ov x} + g(x) + \ov g(\ov x)
\end{equation}
and the K\"ahler potential of the rigid theory will be
$\widehat {\cal K} (x, \ov x) = - i\ov x{\cal F}_x + i x\ov{\cal F}_{\ov x}$. Following Ref.~\cite{FGP2}, 
this is obtained from the formal $\kappa$ expansion, 
\begin{equation} 
f(z) = {1\over4} + \lambda \kappa \tilde\mu\, z + \kappa^2 \Bigl[ i \tilde\mu^2 \widehat F(z) 
+ {1\over4}\tilde\mu^2(\lambda +\ov\lambda)z^2 \Bigr] 
+ {\cal O}(\kappa^3 \tilde\mu^3),
\end{equation}
and the definition
\begin{equation}
{\cal F}(x) = \tilde\mu^2 \, \widehat F(\frac{x}{\tilde\mu}),
\end{equation}
with ${\cal F}_x = \tilde\mu\,\widehat F_z$ and ${\cal F}_{xx} = \widehat F_{zz}$.
The arbitrary complex number $\lambda$ will get a precise value later on.

With the rescaling (\ref{rigidSO14}) of the hypermultiplet scalars, a corresponding rescaling of the Killing vectors,
or equivalently a (first) rescaling of the coupling constants, is needed:
\begin{equation}
g_i = \kappa \tilde\mu \langle b_0 \rangle \, \widetilde g_i \,,
\end{equation}
leading to the scalar potential
\begin{equation}
\label{APTpot1}
\mathscr{V}= \mu^4\, {\text{e}^{\cal K} \over \left(1+ \kappa\mu\widetilde b_0\right)^2 }
\left[ - {1\over\kappa^2\mu^2}\Bigl(\widetilde g_0^2+|c|^2\Bigr) + {1\over\kappa^2\mu^2{\cal K}_{z\ov z}}
\Bigl( \widetilde g_0^2{\cal K}_z {\cal K}_{\ov z}+
| c_z + c {\cal K}_z|^2 \Bigr) \right],
\end{equation}
where $c$ and $c_z$ are expressed in terms of $\widetilde g_i$ (instead of $g_i$),
$c = -i( \widetilde g_1 + \widetilde g_2 z) + \widetilde g_3 f_z$ and 
$c_z = -i \widetilde g_2 + \widetilde g_3 f_{zz}$. 

Before expanding in powers of $\kappa$, we perform a second redefinition of the gauge couplings,
\begin{equation}
\widetilde g_0 = \kappa\tilde\mu \, \widehat g_0 \,, \qquad\quad
\widetilde g_1 = \kappa\tilde\mu \, \widehat g_1 \,, \qquad\quad
\widetilde g_2 = (\kappa\tilde\mu)^2 \, \widehat g_2 \,, \qquad\quad
\widetilde g_3 = \widehat g_3 \,.
\end{equation}
The leading terms in the quantities $c$, $c_z$ and ${\cal K}_z$ appearing in the potential \eqref{APTpot1} 
are then
\begin{eqnarray}
&&c= \left[\widehat g_3\lambda - i\widehat g_1\right]\kappa\tilde\mu + {\cal O}(\kappa^2\tilde\mu^2),
\qquad
c_z = \left[ -i \widehat g_2 + 2\,({\rm Re}\,\lambda)^2\widehat g_3 + i\widehat g_3{\cal F}_{xx} \right]\nonumber
\kappa^2\tilde\mu^2 + {\cal O}(\kappa^2\tilde\mu^2),
\\
&&{\cal K}_z = -2\,{\rm Re}\,\lambda\,\kappa\tilde\mu + {\cal O}(\kappa^2\tilde\mu^2),
\label{cis1}
\end{eqnarray}
and, to leading order in $\kappa$, the potential reads
\begin{eqnarray}
\mathscr{V} &=& \displaystyle  {\tilde\mu^4 \over \widehat{\cal K}_{x\ov x}}\,
\left[ 4({\rm Re}\,\lambda)^2\widehat g_0^2
+ \Bigl| -\widehat g_2 + 2\,\widehat g_1{\rm Re}\,\lambda - 2\, \widehat g_3\, {\rm Re}\,\lambda{\rm Im}\,\lambda
+ \widehat g_3\, {\cal F}_{xx}\Bigr|^2 \right] - C\nonumber
\\
&=&  \displaystyle {1 \over  2\,{\rm Im}\, {\cal F}_{xx}}
\left[ \zeta^2 + | m^2 + M^2 {\cal F}_{xx} |^2 \right] - C
\\
&=&  \displaystyle {1 \over  2\,{\rm Im}\, {\cal F}_{xx}}
\left| m^2 - i\zeta + M^2 {\cal F}_{xx} \right|^2 + \zeta\,M^2 - C,
\nonumber
\end{eqnarray}
where
\begin{equation}
\begin{array}{l}
m^2 = - (\widehat g_2 - 2\,\widehat g_1{\rm Re}\,\lambda + 2\, \widehat g_3 \, {\rm Re}\lambda \,
{\rm Im}\lambda )\tilde\mu^2, \qquad\qquad M^2 = \widehat g_3 \tilde\mu^2,
\crbig
\zeta = 2 \, {\rm Re}\, \lambda\widehat g_0 \,\tilde\mu^2, \qquad\qquad
C = \tilde\mu^4\widehat g_0^2 + \tilde\mu^4|\widehat g_3\lambda -i\widehat g_1|^2 .
\end{array}
\end{equation}
The scalar potential of a globally supersymmetric theory is not expected to have an irrelevant additive constant
and we cancel $\zeta M^2-C$ by choosing
\begin{equation}
\lambda = {1\over\widehat g_3}\, (\widehat g_0 + i \widehat g_1),
\end{equation}
which also implies 
\begin{equation}
\label{cis2}
c = \widehat g_0 \, \kappa\tilde\mu + {\cal O}(\kappa^2\tilde\mu^2) = 
\widetilde g_0 + {\cal O}(\kappa^2\tilde\mu^2).
\end{equation}
This is the leading term in the first condition \eqref{parteq1} for partial breaking (related to the gravitino and 
hyperino shift matrices). 
The shift matrix for canonically normalized (mass$^{\frac{3}{2}}$ dimension) gauginos $\Lambda^i$ becomes
\begin{equation}
\label{shiftrigid}
\delta \, \Lambda^i = {\cal W}_x^{ij}\,\epsilon_j + \cdots, \qquad\qquad
{\cal W}_x^{ij} = {\tilde\mu^2 \over 2 \, {\rm Im}\, {\cal F}_{xx}}
\left( \begin{array}{cc}
4\,\widehat g_0^2 \, \widehat g_3{}^{-1} - c_z  & 0 \\
0 &  c_z 
\end{array}\right) ,
\end{equation}
where ${\cal K}_z$, $c_z$ and $c$ are given in Eqs.~\eqref{cis1} and \eqref{cis2}.\,\footnote{The shift matrices 
$\kappa^2 S^{ij}$ and $N^i{}_A$ vanish when $\kappa\rightarrow0$ and hyperinos decouple from the goldstino.}

Up to here, the analysis has been off-shell only. We now expect that the second condition \eqref{parteq1}
for partial breaking, which indicates that only one gaugino is a goldstino, follows from the 
minimum of the potential, which is at\,\footnote{Metric positivity requires ${\rm Im}\,{\cal F}_{xx}>0$.}
\begin{equation}
\widehat g_3 \langle {\cal F}_{xx}\rangle = \widehat g_2 + 2i\, {\rm Re}\,\lambda \widehat g_0
= \widehat g_2 + 2i\, \widehat g_0^2 / \widehat g_3.
\end{equation}
This vacuum equation is also the leading order term of $\langle c_z \rangle = 0$, the second condition for 
partial breaking \eqref{parteq1}.
And for $\langle c_z \rangle=0$, the gaugino shift matrix \eqref{shiftrigid} has one zero eigenvalue.
At the ground state, the vector multiplet metric is
\begin{equation}
2\, \langle{\rm Im}\,{\cal F}_{xx}\rangle = 4\, \frac{\widehat g_o^2}{\widehat g_3^2},
\end{equation}
and the deformation parameter of the supersymmetry variation in the goldstino direction is then
\begin{equation}
\delta  \Lambda_{\rm goldstino} = M^2 + \cdots = \widehat g_3\,  \tilde\mu^2 + \cdots.
\end{equation}
The coupling constant $\widehat g_3 \tilde\mu^2$ is the magnetic FI term at the origin of the partial breaking. The ${\cal N}=2$ multiplet splits in a massless ${N=1}$ Maxwell, including the goldstino, 
and a chiral ${\cal N}=1$ multiplet with mass
\begin{equation}
M^2_x = \left< {\widehat g_3^6 \mu^4 \over 16\widehat g_0^4}|  {\cal F}_{xxx}  |^2 \right> ,
\end{equation}
as expected from Eq.~\eqref{massscalar}.

In conclusion, we have shown that this ${\cal N}=2$ supergravity theory possesses for all values of $\kappa$
an ${\cal N}=1$ ground state which coincides in the limit $\kappa\rightarrow0$ with the APT lagrangian and its
${\cal N}=1$ vacuum.

\boldmath
\subsection{${\cal N}=0$ Minkowski vacua} \label{secSO4N=0}
\unboldmath

The scalar potential (\ref{potentialAPTSUGRA}) can also be written
\begin{equation}
e^{-1} \mathscr{V}= \frac{1}{\kappa^4\,b_0^2 {\cal K}_{z\ov z} {\cal Y}^2}
\Bigl[ 2\left(g_0^2+|c|^2\right) {\rm Re}\, f_{zz}
+ {\cal Y}|c_z|^2 -\ov c c_z {\cal Y}_{\ov z} - c\ov c_z {\cal Y}_z \Bigr]\,.
\end{equation}
Non-supersymmetric vacua will then follow by solving
\begin{equation}
\label{min}
\partial_z h = \partial_{\ov z} h = h = 0, \qquad\qquad
h(z,\ov z) = 2\left(g_0^2+|c|^2\right) {\rm Re}\, f_{zz}
+ {\cal Y}|c_z|^2 -\ov c c_z {\cal Y}_{\ov z} - c\ov c_z {\cal Y}_z,
\end{equation}
supplemented by stability conditions and the existence of two goldstinos. Since $h$ is real, Eqs.~(\ref{min}) 
give three conditions for the two real components of $z$. Hence, for a given $f(z)$ one expects at least one non-trivial condition on the gauge coupling constants: once $\langle z\rangle$ is fixed by $\partial_zh=0$, the number 
$\langle {\mathscr V}\rangle$ must vanish to avoid runaway in $b_0$.

Since $c_{zz}= g_3\, f_{zzz}$, with the definition \eqref{genericlocalKahler} of ${\cal Y}$, 
it is immediate that $\langle f_{zzz}\rangle = 0$ solves $\partial_z h=0$.
We have already observed that $\langle c_z \rangle = 0$ leads to $h=0$. Hence, 
$\langle f_{zzz}\rangle = \langle c_z \rangle = 0$ is a solution of conditions (\ref{min}).
Since $\langle c_z \rangle =0$ also implies that the gaugino shift matrix has a zero eigenvector, the
determinant of $\langle N^i{}_A\rangle$ should be nonzero in an ${\cal N}=0$ ground state:
\begin{equation}
g_0 \ne \pm\langle c \rangle.
\end{equation}
This happens in the FGP model \cite{FGP} with $f(z)=z$, $g_1=g_2=0$ and then $c= g_3$: all  
$\langle z\rangle$ are stable ground states since $\mathscr{V}\equiv0$, the generic ground state has
${\cal N}=0$ and partial breaking occurs when $g_0 =\pm g_3$. Hence we may think that partial-breaking
solutions are surrounded (in the parameter space of the solutions of a model) by ${\cal N}=0$ solutions.
However, assuming
\begin{equation}
\label{fquadratic}
f(z)=f_0+f_1\, z+f_2\, z^2\,,\qquad\qquad f_{0,1,2}\in\mathbb{C}\,,
\end{equation}
leads to the scalar potential
\begin{equation}
\begin{split}
&{\mathscr{V}=\frac{\cal C}{\kappa^4b_0^2}, \qquad\qquad
{\cal C} = \frac{(g_0^2+|A_1|^2)\text{Re} f_2+|A_2|^2\text{Re} f_0
-\text{Re}f_1\text{Re}(A_1\bar A_2)}{(\text{Re}f_1)^2-4\text{Re}f_0\,\text{Re}f_2}}\,,\crbig
&\qquad\qquad  A_1=g_1+i g_3 f_1\,,\qquad\qquad A_2=g_2+2ig_3f_2 = i\,c_z\,.
\end{split}
\end{equation}
Parameters should be such that the constant ${\cal C}$ vanishes to avoid a runaway in $b_0$.
The choice $c_z=0$ with $g_3f_2\ne0$ leads to ${\cal N}=0$ ground states for arbitrary $\langle z\rangle$
but the supplementary condition for an ${\cal N}=1$ vacuum is never verified.

Working out conditions \eqref{min} leads to two distinct classes of Minkowski vacua:
\begin{enumerate}[label=\roman*.]
\item
All solutions with $\langle f_{zzz}\rangle\neq0$ and $\langle h\rangle=0$ are ${\cal N}=1$ vacua 
already described in Eqs.~\eqref{parteq1}.\footnote{For a proof, see App.~\ref{appproof}.}
\item
Solutions with $\langle f_{zzz}\rangle=0$ and $\langle h\rangle=0$ are generically ${\cal N}=0$ vacua.
\end{enumerate}
Stability of the ${\cal N}=0$ ground states is provided in terms of the mass matrix for the six real scalars $b_u$
and $z$. 
The non-trivial second derivatives of the potential are $\langle\mathscr{V}_{z\ov z}\rangle$ which
vanishes with $\langle f_{zzz}\rangle=0$ and $\langle \mathscr{V}_{zz}\rangle$ which is controlled by the fourth
derivative of $f$. The vacuum is then unstable except if $\langle f^{(n}\rangle=0$ for all $n\geqslant3$ and
this leads us naturally to the choice \eqref{fquadratic}.

\section{Outlook} \label{Sec5}

In summary, our motivation was to classify spontaneous (partial) supersymmetry breaking in the minimal case of 
${\cal N}=2$ supergravity, containing one hypermultiplet and one vector multiplet. The former could describe the 
universal dilaton of type II superstrings compactified on a Calabi--Yau threefold, while the latter should gauge 
together with the ${\cal N}=2$ graviphoton two commuting isometries of the hypermultiplet quaternion-K\"ahler 
manifold, which is necessary in order to obtain a massive ${\cal N}=1$ spin-$3/2$ multiplet. 

The analysis can be done in a general way, since a four-dimensional quaternionic manifold with a two-torus isometry 
can be put in the Calderbank--Pedersen metric form~\cite{CP}. To our surprise, using this approach we found 
a no-go theorem on the existence of ${\cal N}=1$ Minkowski vacua, which would also hold for any number of abelian vector multiplets. This result seems in conflict with the well-known example of the hyperbolic space $SO(4,1)/SO(4)$ \cite{FGP2}. However, we 
proved that the hyperbolic space cannot be written in a Calderbank--Pedersen form, where its torus symmetry lies 
within the three-dimensional abelian subalgebra of $SO(4,1)$. We furthermore showed that it is easy to obtain ${\cal 
N}=1$ vacua of partially broken supersymmetry in AdS space.

Finally, we revisited the hyperbolic space for gaugings within the three-dimensional Abe\-lian subalgebra of 
$SO(4,1)$, while for the scalars of the vector sector we considered
a generic holomorphic prepotential. We worked out the details for a generic gauging
leading to a supergravity theory with potential \eqref{potentialAPTSUGRA} and possessing ${\cal N}=1$ 
or ${\cal N}=0$ Minkowski vacua for all values of $\kappa$. For the ${\cal N}=1$ vacua, we also worked out their off-shell gravity-decoupling limit, and obtained the APT lagrangian \cite{APT}.

Some open questions remain to be answered, which are outside of our present scope.
Regarding the gauged isometries, on the one hand, one may consider gauging isometries of the special-K\"ahler manifold of vector multiplets, 
or (part of) the $SU(2)_R$ $R$-symmetry with 
the compensating hypermultiplet. On the other hand, one may study the effect of more hypermultiplets, for which however explicit and general metrics for quaternion-K\"ahler spaces with isometries are not available. 

\section*{Acknowledgements}

We would like to thank Sergei Alexandrov and Paul Gauduchon for useful discussions and correspondence. The work of IA is supported in part by the 
Swiss National Science Foundation (SNF) and in part by French state funds managed by the Agence Nationale de la Recherche in the context of the LABEX ILP (ANR-11-IDEX-0004-02, ANR-10- LABX-63). 
The work of JPD is not supported by the SNF.
The work of PMP is partly supported by the Agence Nationale pour la Recherche (ANR-16-CE31-0004 contract \textsl{Black-dS-String}).
Finally we acknowledge each-other institutes for hospitality and financial support. 
Part of this work was developed during HEP 2017: Recent Developments in 
High Energy Physics and Cosmology in April 2017 at the University of Ioannina.

\appendix

\section{Four-dimensional quaternionic manifolds with isometries}
\label{appendixquaternionic}

Consider a four-dimensional quaternionic space, described by an Einstein metric with anti-selfdual Weyl
curvature
\begin{equation}
\label{Weylself}
W_{xyrw}+\frac12\,\varepsilon_{xyuv}\,W^{uv}{}_{rw}=0\,,\qquad\qquad R_{uv}=-3\,h_{uv}\,,
\end{equation}
normalized with $R=-12$. This space is endowed with a triplet of $SU(2)$ self-dual
complex structures $J^x_{uv}$, which are covariantly constant modulo an $SU(2)$ one-form connection 
$\omega^x$
\begin{equation}
\label{covJquat}
\nabla_w J^x_{uv}+\varepsilon^{xyz}\,\omega^y_w J^z_{uv}=0\,.
\end{equation}
The complex structures $J^x_{uv}$ are normalized to satisfy: 
\begin{equation}
\label{complexJ}
\left(J^x\right)_u^{\hphantom{u}r}\,\left(J^y\right)_r^{\hphantom{x}v}=
-\delta^{xy}\,\delta_u^v-\varepsilon^{xyz}\,\left(J^z\right)_u^{\hphantom{u}v}\,,\qquad
\left(J^x\right)_u^{\hphantom{u}v}\left(J^x\right)_w^{\hphantom{u}r}=h_{uw}\,g^{vr}-\delta_u^r\, \delta_w^v
+\varepsilon_{uw}^{\hphantom{uw}vr}\,.
\end{equation}
Assume that the quaternionic space has some isometries generated by $\xi_a=\xi_a^u\partial_u$. 
As a consequence of the Bianchi identity for the Riemann tensor,
condition \eqref{Weylself} leads to
\begin{equation}
\nabla_w k_{auv}^+=2P^+_{uvwr}\xi^r_a\,,
\end{equation}
in terms of the (anti)-selfdual covariant derivatives
\begin{equation}
\begin{split}
&k^\pm_{auv}=P^\pm_{uv}{}^{wr}\nabla_w\xi_{ar}\,,\\
&P^\pm_{uv}{}^{wr}=\frac12\left(\delta_{uv}^{wr}\pm\frac12\,\varepsilon_{uv}{}^{wr}\right)\,,\qquad
\delta_{uv}^{wr}:=\frac{1}{2}\left(\delta_u^w\delta_v^r-\delta_u^r\delta_v^w\right)\,.
\end{split}
\end{equation}
These (anti)-selfdual covariant derivatives obey the following identities
\begin{equation}
\begin{array}{ll}
h^{uv}\left(k_{aru}^\pm k_{bwv}^\pm+k_{bru}^\pm k_{awv}^\pm \right)=\frac12\,h_{rw}\,k^\pm_a\cdot k^\pm_b\,,
\qquad &
k^\pm_a\cdot k^\pm_b=h^{rw}h^{uv}k_{aru}^\pm k_{bwv}^\pm\,,\crbig
h^{uv}\left(k_{aru}^\pm k_{bwv}^\mp-k_{bru}^\mp k_{awv}^\pm\right)=0\,,\qquad&
h^{rw}h^{uv}\,k_{aru}^\pm k_{bwv}^\mp=0
\end{array}
\end{equation}
valid for any four-dimensional metric.\footnote{They follow from $SO(4)$ group theory.}

\section{The hyperbolic space and its Calderbank--Pedersen coordinates}
\label{appendixhyperbolic}

\subsection{The hyperbolic space in global and Poincar\'e coordinates}

The $SO(4,1)$ isometry algebra of the hyperbolic space $\text{H}_4$ obtained as $SO(4,1)/SO(4)$ includes six compact $SO(4)$ generators $X_{uv}$ and four noncompact $SO(4,1)/SO(4)$ generators $Y_u= X_{u5}$, with 
$\eta_{55}=-1=-\eta_{uu}$.
It has a three-dimensional abelian subalgebra related to noncompactness. In the standard 
notation or the $SO(4,1)$ algebra,\footnote{The same would hold for $X_{i4}-X_{i5}$.}
\begin{equation}
[ X_{i4} + X_{i5} , X_{j4} + X_{j5} ] = -(\eta_{44}+\eta_{55}) X_{ij} =0, \qquad\qquad i,j=1,2,3.
\end{equation}
The commuting $r_i = X_{i4}+Y_i$ form a vector of $SO(3) \subset SO(4)$.

We can describe  $\text{H}_4$ in global coordinates. 
In this set of coordinates the $SO(4)$ acts linearly and the line element takes the form
\begin{equation}
\label{hyperbolic.linear}
\text{d}s^2=\frac{4\,\text{d}x^u\text{d} x^u}{\left(1- x^v x^v\right)^2}\,,\qquad x^u=(x_1,x_2,x_3,x_4)\,.
\end{equation}
Its ten isometry generators are:
\begin{equation}
\begin{array}{lrcl}
\text{Generators of}\,\, SO(4):&\quad X_{uv} &=&x^v\partial_u-x^u\partial_v\,,\crbig
\text{Generators of}\,\, SO(4,1)/SO(4):&\quad Y_u &=& \displaystyle
\frac{4+x^v x^v}{4}\partial_u-\frac12\,x^u x^v\partial_v\,,
\end{array}
\end{equation}
where $\partial_u=\frac{\partial}{\partial x^u}$. The three-dimensional abelian subalgebra is
generated by
\begin{equation}
r_1=X_{14}+Y_1\,,\qquad r_2=X_{24}+Y_2\,,\qquad r_3=X_{34}+Y_3\,.
\end{equation}
In these coordinates, the $SO(4)$ generators act as simple linear variations but the action of the commuting $r_i$ is more involved.
The curvature is directly related to the $SO(4,1)$-invariant
quantity $1- x^v x^v$, and these coordinates are then convenient for describing the (flat or rigid)  gravity-decoupling
limit.

An alternative coordinate system is  the Poincar\'e patch
 $b^u=(b_0,b_1,b_2,b_3)$. 
The metric takes the form \eqref{metricSO}
\begin{equation}
\text{d}s^2=\frac{\text{d}b_0^2+\text{d}b_1^2+\text{d}b_3^2+\text{d}b_3^2}{b_0^2}\,.
\end{equation}
In these coordinates, the generators of the three-dimensional abelian subalgebra act as translations of
$b^i$:
\begin{equation}
r_i = {\partial\over\partial b^i}.
\end{equation}

The two sets of coordinates are related by
\begin{equation}
b_0=\frac{4-x^ux^u}{4(1+x^4)+x^ux^u}\,,\qquad
b_i=\frac{4x^i}{4(1+x^4)+x^ux^u}\,,\qquad
i=1,2,3\,.
\end{equation}

\subsection{Calderbank--Pedersen coordinates}

The isometry algebra $SO(4,1)$ admits a variety of pairs of commuting generators and for each pair,
according to Ref.~\cite{CP}, there should exist CP coordinates $\rho$, $\eta$, $\varphi$, $\psi$. Examples 
of (inequivalent) pairs are:
\begin{enumerate}[label=\roman*.]
\item
A pair of isometries in the three-dimensional abelian subalgebra, for instance $r_1$ and $r_2$.
\item
The Cartan subalgebra of $SO(4)$, chosen as the compact generators $X_{12}$ and $X_{34}$,
or a compact and a non compact $SO(4,1)$ generator, like $X_{12}$ and $Y_4=X_{45}$.
\item
A compact generator of $SO(4)$ and one of the $r_i$'s, for instance $X_{23}$ and $r_1$.
\end{enumerate}
In each case, there are equivalent choices obtained by either $SO(4)$ or $SO(3)$ rotations. The case (\romannumeral3) is only one example of pairing one $SO(4)$ generator with any generator in the $SO(2,1)$ algebra 
commuting with it.

We have shown in \S \ref{H2CP} that CP coordinates do not exist for the case (\romannumeral1). We here show how CP 
coordinates can be derived for cases (\romannumeral2) and (\romannumeral3).

\paragraph{Case (\romannumeral2) -- the Cartan algebra of {\boldmath$SO(4)$}} This is easily analyzed in coordinates
where the $SO(4)$ has a linear action, {\it i.e.} coordinates \eqref{hyperbolic.linear}.
The commuting (compact) Killing vectors are rotations in planes $(12)$ and $(34)$
\begin{equation}
\xi_1=x_2\partial_{x_1}-x_1\partial_{x_2}\,,\qquad \xi_2=x_4\partial_{x_3}-x_3\partial_{x_4}\,.  
\end{equation}
We next identify $\xi_{1,2}$ with the Killing vectors of the CP metric or with linear combinations of them, 
and use the identity \eqref{QKtoCP} for recognizing  the change of coordinates. There are actually several
possibilities and we focus on two cases, following Ref.~\cite{CP}.
\begin{itemize}
\item
The identification $(\partial_\varphi,\partial_\psi)=(\xi_1,\xi_2)$ leads to
\begin{equation}
\begin{array}{l} \displaystyle
\left(r_1+i r_2\right)^2=\frac{\eta+1-i\rho}{\eta-1-i\rho}\,, \qquad\qquad
r_1^2=x_1^2+x_2^2\,,\qquad r_2^2=x_3^2+x_4^2\,,
\crbig \displaystyle
F(\rho,\eta)=\frac{1}{2\sqrt{\rho}}\left(\sqrt{\rho^2+(\eta+1)^2}-\sqrt{\rho^2+(\eta-1)^2}\right).
\end{array}
\end{equation}
\item 
Choosing instead $(\partial_\varphi,\partial_\psi)=(\xi_1+\xi_2,\xi_1-\xi_2)$, we obtain
\begin{equation}
\rho=2r_1r_2\,,\qquad\qquad \eta=r_1^2-r_2^2\,,\qquad\qquad 
F(\rho,\eta)=\frac{1}{2\sqrt{\rho}}\left(\sqrt{\rho^2+\eta^2}-1\right).
\end{equation}
\end{itemize}
Choosing instead $\xi_1=X_{12}$, $\xi_2=Y_4=X_{45}$ and using coordinates $b^u$, the Killing vectors are
\begin{equation}
\xi_1=b_2\,\partial_{b_1}-b_1\,\partial_{b_2}\,, \qquad\qquad \xi_2= -b_0\,\partial_{b_0}-b_1\,\partial_{b_1}
-b_2\,\partial_{b_2}-b_3\,\partial_{b_3}\,.
\end{equation}
Working as above we obtain:
\begin{equation}
\begin{split}
&\rho= r \frac{\sqrt{b_0^2+r^2+b_3^2}}{r^2+b_3^2}\,,\qquad\quad
\eta=-\frac{b_0b_3}{r^2+b_3^2}\,,\quad\qquad r^2=b_1^2+b_2^2\,,\\
&F=\frac{2^{1/4}\eta}{\sqrt{\rho}\left(t^2+(1-\rho^2-\eta^2)t-2\eta^2\right)^{1/4}}\,,\qquad 
t=\sqrt{(\rho+1)^2+\eta^2}\sqrt{(\rho-1)^2+\eta^2}\,.
\end{split}
\end{equation}
In any case, since CP coordinates exist, gauging these isometries does not lead to partial breaking.

\paragraph{Case (\romannumeral3) -- {\boldmath$r_1$} and {\boldmath$X_{23}$}} This case is more easily analyzed in coordinates $b^u$ where $r_1$ is a translation of $b^1$:
\begin{equation}
\xi_1=\partial_{b_1}\,,\qquad\qquad \xi_2=b_3\partial_{b_2}-b_2\partial_{b_3}\,.
\end{equation}
We again consider two cases:
\begin{itemize}
\item 
With $(\partial_\varphi,\partial_\psi)=(\xi_1,\xi_2)$, we obtain
\begin{equation}
\rho=\frac{r}{b_0^2+r^2}\,,\quad \eta=\frac{b_0}{b_0^2+r^2}\,,\quad 
F(\rho,\eta) =\frac{\eta}{\sqrt{\rho(\rho^2+\eta^2)}}\,,
\end{equation}
where $r^2=b_2^2+b_3^2$.
\item 
The choice $(\partial_\varphi,\partial_\psi)=(\xi_2,\xi_1)$ leads to 
\begin{equation}
\rho=r\,,\qquad \eta=b_0\,,\qquad F(\rho,\eta) =\frac{\eta}{\sqrt{\rho}}\,.
\end{equation}
\end{itemize}
Again, CP coordinates exist and these isometries do not induce partial breaking.

Finally, for completeness and out of curiosity, we present the CP form of the sphere $SO(5)/SO(4)$ 
(which cannot describe a hypermultiplet),
where all pairs of commuting isometries are equivalent to $X_{12}$ and $X_{34}$. In coordinates
where
\begin{equation}
{\rm d}s^2 ={4\, {\rm d}x^u {\rm d}x^u \over (1+x^vx^v)^2} \,, \qquad
\xi_1= x_2\partial_{x_1} - x_1\partial_{x_2} \,,\qquad
\xi_2= x_4\partial_{x_3} - x_3\partial_{x_4} \,,
\end{equation}
we again consider two choices of identification \cite{CP}:
\begin{itemize}
 \item 
Now $(\partial_\varphi,\partial_\psi)=(\xi_1,\xi_2)$ leads to 
\begin{equation}
\begin{split}
&\left(r_1+i r_2\right)^2=\frac{\eta+1-i\rho}{\eta-1-i\rho}\,,\\
&F(\rho,\eta) =\frac{1}{2\sqrt{\rho}}\left(\sqrt{\rho^2+(\eta+1)^2}+\sqrt{\rho^2+(\eta-1)^2}\right)\,,\\
&r_1^2=x_1^2+x_2^2\,,\qquad r_2^2=x_3^2+x_4^2\,.
\end{split}
\end{equation}
 \item 
For $(\partial_\varphi,\partial_\psi)=(\xi_1+\xi_2,\xi_1-\xi_2)$ we obtain
\begin{equation}
\rho=2\,r_1r_2\,,\qquad \eta=r_1^2-r_2^2\,,\qquad
F(\rho,\eta)=\frac{1}{2\sqrt{\rho}}\left(\sqrt{\rho^2+\eta^2}+1\right).
\end{equation}
\end{itemize}

\section{Ward transformation} \label{secWard}

Assume that we have a solution $V(\rho,\eta)$ of the equation
\begin{equation}
\label{Veq}
{1\over\rho}\left(\rho V_\rho\right)_\rho + V_{\eta\eta} = 0,
\end{equation}
where indices denote derivatives with respect to $\eta$ or $\rho$.
A further derivative with respect to $\rho$ leads to
\begin{equation}
{\partial^2 F\over\partial\rho^2} + {\partial^2F\over\partial\eta^2} = {3F\over4\rho^2}
\qquad\makebox{with}\qquad F(\rho,\eta) =\sqrt\rho\,V_\rho,
\end{equation}
and $F(\rho,\eta)$ generates via Eq.~(\ref{CPMetric}) a quaternion-K\"ahler metric in Calderbank--Pedersen coordinates. Coordinates $(\rho,\eta)$ can be traded for $(X,Z)$ by a double Legendre transformation:
\begin{equation}
V(\rho,\eta) - X\eta -2\,Z\ln\rho = - K(X,Z).
\end{equation}
This implies firstly
\begin{equation}
\rho V_\rho = 2\,Z, \qquad V_\eta=X, \qquad
\eta= K_X, \qquad 2\,\ln\rho=K_Z .
\end{equation}
Secondly
\begin{equation}
{\partial Z\over\partial\rho} = {1\over2}\left(\rho V_\rho\right)_\rho, \qquad
{\partial Z\over\partial\eta} = {\rho\over2}V_{\rho\eta}, \qquad
{\partial X\over\partial\rho} =V_{\rho\eta}, \qquad
{\partial X\over\partial\eta} = V_{\eta\eta},
\end{equation}
and
\begin{equation}
{\partial\rho\over\partial X} = {\rho\over2}K_{XZ}, \qquad
{\partial\rho\over\partial Z} = {\rho\over2}K_{ZZ}, \qquad
{\partial\eta\over\partial X} = K_{XX}, \qquad
{\partial\eta\over\partial Z} = K_{XZ}.
\end{equation}
As usual, ${\partial x^i\over\partial x^j} = \delta^i_j$ for each set of coordinates delivers the relations between 
the second derivatives of $V$ and $K$. Using then Eq.~(\ref{Veq}), the relevant equation appears to be
\begin{equation}
\label{Keq}
K_{XX} + {\rho^2\over4}\, K_{ZZ} = 0
\end{equation}
as the ``Legendre partner'' of Eq.~(\ref{Veq}).
Define finally
\begin{equation}
\Psi(X,Z) = \ln\left({1\over4}\rho^2\right), \qquad\qquad \text{e}^\Psi= {1\over4} \, \rho^2 = {1\over4}\, \text{e}^{K_Z}.
\end{equation}
The relations induced by the Legendre transformation and Eq.~(\ref{Keq})
lead to Toda equation for $\Psi$:
\begin{equation}
\Psi_{XX} + \left(\text{e}^\Psi\right)_{ZZ}=0.
\end{equation}
This procedure has been elaborated by Ward in Ref.~\cite{Ward} and used in the derivation of the 
CP metric \cite{CP}.
It allows in particular to find CP coordinates for a quaternion-K\"ahler metric with two isometries expressed in 
PT coordinates, for a given Toda solution $\Psi$.

The case where $\Psi $ is a constant is clearly excluded.

\section{A proof} \label{appproof}

In \S~\ref{secSO4N=0} on ${\cal N}=0$ vacua of the $SO(4,1)/SO(4)$ model, we claim that \emph{all solutions of 
$\partial_z h = h = 0$ with $f_{zzz}\ne 0$ are ${\cal N}=1$ vacua}.\,\footnote{In order to avoid cluttering in the formulas, we systematically omit $\langle\ldots\rangle$ 
in this appendix.} We give here a proof of this statement.

Recall that, for a given prepotential function $f(z)$,
\begin{equation}
{\cal Y} = 2\left(f+\ov f\right) - (z-\ov z)\left(f_z - \ov f_{\ov z}\right), \quad\quad
c = -i(g_1+g_2 z) + g_3 f_z, \quad\quad g_3\ne0.
\end{equation}
Starting with
\begin{equation}
\begin{array}{rcl}
h &=& \left(g_0^2+|c|^2\right) \left(f_{zz} + \ov f_{\ov{zz}}\right)
+ {\cal Y}|c_z|^2 -\ov c c_z {\cal Y}_{\ov z} - c\ov c_{\ov z} {\cal Y}_z \,,
\crbig
\partial_z h &=& f_{zzz} \Bigl[ g_0^2 + c\ov c + g_3 \,\ov c_{\ov z}\, {\cal Y} - g_3\,\ov c\, {\cal Y}_{\ov z} 
+(z-\ov z)c \ov c_{\ov z} \Bigr]
\crbig
&=& f_{zzz}\Bigl[  (g_0+\ov c)(g_0-\ov c) + \ov c_{\ov z} \left[ g_3{\cal Y} + (z-\ov z)(c-\ov c)\right]\Bigr]
\end{array}
\end{equation}
and assuming $f_{zzz}\ne0$, one finds the factorization:
\begin{equation}
c_z \, (\partial_z\, h) \, f_{zzz}{}^{-1} + \ov c_{\ov z} \, (\partial_{\ov z}\, h) \, \ov f_{\ov{zzz}}{}^{-1} - g_3\, h
= c_z \, \ov c_{\ov z} \, \Bigl[ g_3\,{\cal Y} +  (z-\ov z)(c-\ov c) \Bigr].
\end{equation}
This quantity should vanish for a solution of $\partial_z h = h = 0$. The solutions are either
$c_z=0$ or $g_3\,{\cal Y} +  (z-\ov z)(c-\ov c)=0$ and in both cases $\partial_z\, h=0$
requires $c = \pm g_0$.
\begin{itemize}
\item If $c_z=0$ and $c=\pm g_0$ the vacuum has ${\cal N}=1$ supersymmetry: the two conditions for partial breaking
\eqref{parteq1} are fulfilled.
\item If $c_z\ne0$, the vacuum state would be at $g_3\,{\cal Y} +  (z-\ov z)(c-\ov c)=0$ and $c = \pm g_0=\ov c$. 
Then ${\cal Y}= \text{e}^{-{\cal K}}=0$,
which is excluded.
\end{itemize}
Hence, \emph{Minkowski ${\cal N}=0$ vacua with $f_{zzz}\ne0$ do not exist}.


\end{document}